\documentclass[12pt]{iopart}

\usepackage{iopams}
\usepackage{graphicx}
\usepackage[usenames,dvipsnames]{color}

\begin{document}

\title[Euclidean random matrices for waves in random media]{Eigenvalue distributions of large Euclidean random matrices for waves in random media}

\author{S E Skipetrov and A Goetschy}

\address{Laboratoire de Physique et Mod\'{e}lisation des Milieux Condens\'{e}s, Universit\'{e} Joseph Fourier and CNRS UMR 5493, B.P. 166, 25 rue des Martyrs,
Maison des Magist\`{e}res, 38042 Grenoble Cedex 09, France}

\ead{Sergey.Skipetrov@grenoble.cnrs.fr}

\begin{abstract}
We study probability distributions of eigenvalues of Hermitian and non-Hermitian Euclidean random matrices that are typically encountered in the problems of wave propagation in random media.
\end{abstract}


\section{Introduction}
\label{secintro}

Random matrix theory is a powerful tool of statistical physics \cite{mehta91} with important applications in the field of quantum and wave transport in random media \cite{beenakker97,fyodorov97,guhr98}. A special class of random matrices are Euclidean random matrices with elements $F_{ij}$ defined with the help of some function $f(\mathbf{r}_i, \mathbf{r}_j)$: $F_{ij} = f(\mathbf{r}_i, \mathbf{r}_j)$. Here $\mathbf{r}_i$ ($i = 1, \ldots N$) are randomly chosen points in the Euclidean space \cite{mezard99,parisi06}.
Euclidean random matrices appear in various physical contexts and were previously considered to interpret the `boson peak' in supercooled liquids \cite{grigera03} or to study slow relaxation in glasses and scalar phonon localization \cite{amir10}, to cite a few recent examples. The purpose of this paper is to study eigenvalue distributions of certain large Euclidean random matrices that appear in problems of wave propagation in random media. Because in the simplest case of scalar waves the propagation is described by a scalar wave equation, the function $f$ that will be of interest to us is the Green's function $G(\mathbf{r}_i, \mathbf{r}_j)$ of the Helmholtz equation
\begin{eqnarray}
\left( \nabla^2 + k_0^2 + \rmi \eta \right) G(\mathbf{r}_i, \mathbf{r}_j) = -\frac{4 \pi}{k_0} \delta(\mathbf{r}_i - \mathbf{r}_j),
\label{helmholtz}
\end{eqnarray}
where $\eta$ is a positive infinitesimal.
It is easy to check that
$G(\mathbf{r}_i, \mathbf{r}_j) = \exp(\rmi k_0 |\mathbf{r}_i - \mathbf{r}_j|)/k_0 |\mathbf{r}_i - \mathbf{r}_j|$ \footnote{We restrict ourselves to three-dimensional space in this paper.}.

Statistical properties of the ensemble of matrices ${\hat G}$ with elements $G_{ij} = (1 - \delta_{ij}) G(\mathbf{r}_i, \mathbf{r}_j)$ for $N \gg 1$ are of primary importance in the context of Anderson localization of electromagnetic \cite{rusek96,rusek00,pinheiro04} and matter \cite{antezza10} waves. The same matrices appear in the studies of collective spontaneous emission in dense atomic systems \cite{ernst69,svid08,scully09,svid09,svid10}. The interplay between Anderson localization and Dicke superradiance can also be described by this ensemble of matrices \cite{akker08} and properties of their eigenvalues are important for understanding of random lasers \cite{pinheiro06,goetschy10} and dynamic instabilities in nonlinear random media \cite{gremaud10}. Meanwhile, the matrix ${\hat G}$ is non-Hermitian, its eigenvalues are complex and their probability distribution is difficult to access. This is why in several works dealing with superradiance \cite{svid08,svid09,svid10,akker08} the imaginary part of ${\hat G}$, a matrix with elements $\sin(k_0 |\mathbf{r}_i - \mathbf{r}_j|)/k_0 |\mathbf{r}_i - \mathbf{r}_j|$, was considered. This real symmetric matrix is much easier to study and in many situations it still contains some of the important aspects of the full problem. Similarly, the real part of ${\hat G}$, a matrix with elements $\cos(k_0 |\mathbf{r}_i - \mathbf{r}_j|)/k_0 |\mathbf{r}_i - \mathbf{r}_j|$, is relevant for understanding the collective Lamb shifts in dense atomic systems \cite{scully09,svid10}.

Despite the importance of the three matrices ${\hat G}$, ${\hat S} = \mathrm{Im} {\hat G}$ and ${\hat C} = \mathrm{Re} {\hat G}$ introduced above little is known about statistical properties of their eigenvalues. In the general case, the eigenvalue distribution of ${\hat G}$ was studied only numerically \cite{rusek96,rusek00,pinheiro04}.
Some analytic results are available in the limit of high density of points $\mathbf{r}_i$ inside a sphere: $\rho = N/V \rightarrow \infty$ \cite{svid08,scully09,svid09,svid10}, when the summation in the eigenvalue equation $\sum_{j} G_{ij} \psi_j = \lambda \psi_i$ can be replaced by integration. The purpose of this paper is to partially fill this gap by considering eigenvalue distributions of the three matrices above at finite densities $\rho$,
with the distances between neighboring points $\mathbf{r}_i$ that are larger than, comparable, or smaller than the wavelength $\lambda_0 = 2\pi/k_0$. This situation is of particular importance in the context of wave propagation in random media because in order to observe phenomena due to scattering of waves on the heterogeneities of the medium, the density of scattering centers (or scatterers) should be neither too low (in this case the scattering is negligible), nor too high (in this case the medium responds as an effective homogeneous medium).
One of the possible experimental realizations of a strongly scattering system is a cloud of cold atoms in which propagation of quasi-resonant light (wavelength $\lambda_0$) is studied \footnote{Light is a vector wave but here we restrict ourselves to a scalar approximation.}. Nowadays such clouds are routinely created at densities $\rho \lambda_0^3 \ll 1$, allowing observation of interesting phenomena due to the multiple scattering of light \cite{labeyrie99,labeyrie03}. This justifies the importance of properly understanding the low-density regime.
However, the most interesting phenomena for waves in an ensemble of point-like scattering centers are known to take place at densities $\rho \lambda_0^3 \gtrsim 1$, when interference effects become important, eventually leading to Anderson localization (see, e.g., \cite{anderson58,lagendijk09,alspecial10} and references therein). Our results may be useful for understanding Anderson localization and its interplay with other collective phenomena (such as Dicke superradiance) \cite{akker08}.

\newpage
\section{Summary of main results}
\label{sum}

\noindent
Before presenting the details of calculations, let us list our main results:

\begin{itemize}

\item
A general framework is developed to deal with Hermitian Euclidean matrices (\sref{secframe}). We show how the theory of asymptotically free random variables can be applied in this context.

\item
The approach developed in \sref{secframe} is applied to study the probability distribution $p(\lambda)$ of real eigenvalues $\lambda$ of the real symmetric random matrix ${\hat S}$ corresponding to $N$ points in a box of side $L$ (\sref{secsinc}). We show that when $\beta = 2.8 N/(k_0 L)^2 < 1$, $p(\lambda)$ is given by the famous Marchenko-Pastur with $\beta = \mathrm{var} \lambda$ as the only parameter. For $\beta > 1$, the Marchenko-Pastur law does not apply anymore.

\item
The probability distribution $p(\lambda)$ of real eigenvalues $\lambda$ of the real symmetric random matrix ${\hat C}$ is studied (\sref{seccosc}). We show that $p(\lambda)$ depends on two parameters: $\beta$ and the number of points per wavelength cube $\rho \lambda_0^3$. Analytic results are in agreement with numerical simulations for $\rho \lambda_0^3 \lesssim 30$ and any $\beta$. In the low-density limit $\rho \lambda_0^3 \ll 1$, $p(\lambda)$ exhibits a transition from the Wigner semi-circle law for $\beta \ll 1$ to the Cauchy distribution for $\beta \gg 1$.

\item
As the first example of non-Hermitian Euclidean matrices, in \sref{secfree} we study the complex symmetric matrix
${\hat X } = {\hat C} + \rmi ({\hat S}' - {\hat \mathbb{I}})$, where two different and independent sets of points $\{ \mathbf{r}_i \}$ and $\{ \mathbf{r}_i' \}$ are used to define the matrices ${\hat C}$ and ${\hat S}'$. For $\beta < 1$, the probability distribution of complex eigenvalues $\lambda$ of ${\hat X}$ is obtained by combining the results for ${\hat S}$ and ${\hat C}$ obtained in \sref{secsinc} and \sref{seccosc}, respectively, in the framework of the theory of free random variables. The domain of existence of eigenvalues of ${\hat X}$ undergoes a transformation from a circular to a triangular shape as $\beta$ increases from 0 to 1. For $\beta \gg 1$, numerical simulations show that the support of the distribution on the complex plane takes an `inverted T' shape.

\item
The non-Hermitian matrix ${\hat G} = {\hat C} + \rmi ({\hat S} - {\hat \mathbb{I}})$ with elements given by the Green's function of the Helmholtz equation \eref{helmholtz} is studied in \sref{secexpc} by means of extensive numerical simulations. We find that at low density $\rho \lambda_0^3 \lesssim 30$ and for $\beta \ll 1$ the domain of existence of eigenvalues of ${\hat G}$ on the complex plane coincides with that of ${\hat X}$ and is given by a circle of radius $\sqrt{2 \beta}$ centered at $(0, \frac12 \beta)$. At larger $\beta$, the domain remains approximately a circle with the same center but a larger radius $R \approx \sqrt{2 \beta + (\frac12 \beta)^2}$. When the density $\rho \lambda_0^3$ reaches a critical values of approximately 30, a `hole' opens in the eigenvalue distribution that otherwise still keeps its circular shape.

\item
The numerically evaluated marginal distributions of real and imaginary parts of the eigenvalues $\lambda$ of the matrix ${\hat G}$ roughly follow the laws obtained for the eigenvalues of the matrices ${\hat C}$ (for $\rho \lambda_0^3 \lesssim 30$) and ${\hat S}$ (for $\frac12 \beta < 1$), respectively. For $\frac12 \beta > 1$, the distribution of $\Gamma = \mathrm{Im} \lambda + 1$ approaches the $1/\Gamma$ law.

\item
The mean minimum value of $\mathrm{Im} \lambda$ is approximately given by $\langle \min(\mathrm{Im} \lambda) \rangle \simeq -1 + 2.3/(N \times \rho \lambda_0^3)^{2/3}$ for $\rho \lambda_0^3 \lesssim 10$ and decays faster at higher densities. The mean maximum value of $\mathrm{Im} \lambda$ is roughly $\langle \max(\mathrm{Im} \lambda) \rangle \simeq \frac12 \beta + R$.

\end{itemize}

The above mathematical results have important applications in a number of physical problems of contemporary interest, as discussed in \sref{appl}. In particular, they provide an additional insight into the cooperative spontaneous emission of large atomic clouds (\sref{coop}), Anderson localization (\sref{loca}), and random lasing (\sref{lasers}).

\section{General framework}
\label{secframe}

Consider a singly-connected three-dimensional region of space $V$. Let $\{ \psi_m(\mathbf{r}) \}$ be an orthonormal basis in $V$, such that
\begin{eqnarray}
\int\limits_V \rmd^3 \mathbf{r}\; \psi_m(\mathbf{r})
\psi_n^*(\mathbf{r}) = \delta_{mn}.
\label{basis}
\end{eqnarray}
We will now show that an arbitrary $N \times N$ Euclidean random matrix
${\hat F}$ with elements
\begin{eqnarray}
F_{ij} = f(\mathbf{r}_i, \mathbf{r}_j),
\;\;\; i,j = 1, \ldots N,
\label{fij}
\end{eqnarray}
where $f$ is a sufficiently well-behaved function of $\mathbf{r}_i, \mathbf{r}_j \in V$, can be represented as
\begin{eqnarray}
{\hat F} = {\hat H} {\hat T} {\hat H}^{\dagger}.
\label{hth}
\end{eqnarray}
Here ${\hat H}$ is a $N \times M$ matrix with elements
\begin{eqnarray}
H_{im} = \sqrt{\frac{V}{N}} \psi_m(\mathbf{r}_i).
\label{him}
\end{eqnarray}
We use $V$ to denote the considered three-dimensional region of space as well as its volume, ${\hat T}$ is a $M \times M$ matrix to be defined below, and the dagger `$\dagger$' denotes Hermitian conjugation.
The size $M$ of the matrix ${\hat T}$ can be arbitrary and, in fact, $M$ will be infinite for the majority of functions $f(\mathbf{r}_i, \mathbf{r}_j)$.

To establish \eref{hth}, we write the $ij$'th element of the matrix ${\hat F}$ explicitly as
\begin{eqnarray}
F_{ij} = \frac{V}{N} \sum\limits_{m, n}
T_{mn} \psi_m(\mathbf{r}_i) \psi_n^*(\mathbf{r}_j),
\label{fij2}
\end{eqnarray}
where we used \eref{him} and the definition of matrix multiplication. Multiplying this equation by
$\psi_{m'}^*(\mathbf{r}_i) \psi_{n'}(\mathbf{r}_j)$, integrating
over $\mathbf{r}_i$ and $\mathbf{r}_j$, and using the orthogonality of the basis functions $\psi_m(\mathbf{r})$, we readily obtain
\begin{eqnarray}
T_{mn} = \frac{N}{V}
\int\limits_V \rmd^3 \mathbf{r}_i \int\limits_V \rmd^3 \mathbf{r}_j\;
f(\mathbf{r}_i, \mathbf{r}_j)
\psi_m^*(\mathbf{r}_i) \psi_n(\mathbf{r}_j).
\label{tmn}
\end{eqnarray}
It is easy to check that with the elements $T_{mn}$ of ${\hat T}$ defined by \eref{tmn}, \eref{hth} is indeed obeyed.

When the points $\{\mathbf{r}_i \}$ are chosen inside $V$ randomly, ${\hat F}$ and ${\hat H}$ become random matrices, whereas ${\hat T}$ is always a non-random matrix independent of $\{\mathbf{r}_i \}$ and determined uniquely by the function $f$, the region $V$, and the choice of the orthonormal basis $\{ \psi_m(\mathbf{r}) \}$.
We will limit our consideration to the case when the spatial integral of any basis function $\psi_m(\mathbf{r})$ that contributes to \eref{fij2} vanishes \footnote{This restricts the class of functions $f(\mathbf{r}_i, \mathbf{r}_j)$ to which our analysis applies but will be sufficient for us here.}:
\begin{eqnarray}
\int\limits_{V} \rmd^3 \mathbf{r}\; \psi_m(\mathbf{r}) = 0.
\label{intvan}
\end{eqnarray}
The elements $H_{im}$ of ${\hat H}$ are then independent random variables having zero means and variances equal to $1/N$:
\begin{eqnarray}
\langle H_{im} \rangle &=& \frac{1}{V} \int\limits_{V}\rmd^3 \mathbf{r}_i\; \sqrt{\frac{V}{N}} \psi_m(\mathbf{r}_i) = 0,
\label{matrixh1}
\\
\langle H_{im} H_{jn}^* \rangle &=& \frac{1}{V^2} \int\limits_{V} \rmd^3 \mathbf{r}_i \int\limits_{V} \rmd^3 \mathbf{r}_j\; \frac{V}{N} \psi_m(\mathbf{r}_i) \psi_n^*(\mathbf{r}_j)
\nonumber \\
&=& \langle H_{im} \rangle \langle H_{jn}^* \rangle = 0,
\;\;\; i \ne j,
\\
\langle H_{im} H_{in}^* \rangle &=& \frac{1}{V} \int\limits_{V} \rmd^3 \mathbf{r}_i\; \frac{V}{N} \psi_m(\mathbf{r}_i) \psi_n^*(\mathbf{r}_i) = \frac{\delta_{mn}}{N}.
\label{matrixh3}
\end{eqnarray}
The representation \eref{hth} is very useful because it can be dealt with using the powerful mathematical arsenal of the so-called free random variable theory \cite{voiculescu92,tulino04,janik97}. Without going into details, we remind the reader that for random matrices, the notion of asymptotic freeness \cite{voiculescu92} is equivalent to the notion of statistical independence that we are familiar with for random variables. Three fundamental objects of the free random variable theory, defined for any Hermitian matrix $\hat F$, will be useful for us in this paper: the usual Green's function
\begin{eqnarray}
{\cal G}(z)=\frac{1}{N}\left\langle\mathrm{Tr}\frac{1}{z-{\hat F}}\right\rangle,
\label{green}
\end{eqnarray}
the Blue function $B(z)$ equal to the functional inverse of ${\cal G}(z)$:
\begin{eqnarray}
B[{\cal G}(z)] = z,
\label{blue}
\end{eqnarray}
and the  $S$-transform of the probability distribution of eigenvalues defined through an auxiliary function $\chi(z)$:
\numparts
\begin{eqnarray}
&&S(z) =
\frac{1 + z}{z} \chi(z),
\label{sg1}
\\
&&\frac{1}{\chi(z)} {\cal G}\left[ \frac{1}{\chi(z)} \right] - 1 = z.
\label{sg2}
\end{eqnarray}
\endnumparts
If two Hermitian random matrices ${\hat A}$ and ${\hat B}$ are asymptotically free, the Blue function $B_{\hat C}(z)$ of their sum ${\hat C} = {\hat A} + {\hat B}$ is equal to the sum of individual Blue functions $B_{\hat A}(z)$ and $B_{\hat B}(z)$, minus $1/z$. The $S$-transform of the matrix product ${\hat C} = {\hat A} {\hat B}$ can be found by multiplying the individual $S$-transforms of ${\hat A}$ and ${\hat B}$. Once the Blue function or the $S$-transform corresponding to the random matrix ${\hat C}$ are found, its Green's function ${\cal G}(z)$ can be calculated either from \eref{blue} or from \eref{sg1} and \eref{sg2}. The probability density of the eigenvalues $\lambda$ of ${\hat C}$ is then determined in the usual way:
\begin{eqnarray}
p(\lambda) = -\frac{1}{\pi}
\lim\limits_{\epsilon \rightarrow 0^+}
\mathrm{Im} {\cal G}(\lambda + \rmi \epsilon).
\label{pimg}
\end{eqnarray}

The functions ${\cal G}(z)$, $B(z)$ and $S(z)$ all contain the same full information about the statistical distribution of eigenvalues $\lambda$ as $p(\lambda)$. The Green's function can be represented as a series with coefficients in front of consecutive powers of $1/z$ equal to statistical moments of $\lambda$:
${\cal G}(z) = \sum_{n=0}^{\infty} \langle \lambda^n \rangle /z^{n+1}$. We have, therefore,
\begin{eqnarray}
\langle \lambda^n \rangle =
\left. \frac{1}{(n+1)!} \frac{d^{n+1} {\cal G}(z)}{d(1/z)^{n+1}}
\right|_{z \rightarrow \infty},
\label{momentg}
\end{eqnarray}
where $z$ is assumed real. Using this equation and \eref{blue} we readily derive an expression for $\langle \lambda^n \rangle$ in terms of $B(z)$:
\begin{eqnarray}
\langle \lambda^n \rangle =
\left. \frac{1}{(n+1)!}
\left[ -\frac{B^2(z)}{B'(z)} \frac{d}{dz} \right]^n
\left[ -\frac{B^2(z)}{B'(z)} \right]
\right|_{z \rightarrow 0},
\label{momentb}
\end{eqnarray}
where $B'(z) = dB(z)/dz$.
If we introduce the $R$-transform ${\cal R}(z) = B(z) - 1/z$ \cite{tulino04}, the average eigenvalue and the variance become
$\langle \lambda \rangle = {\cal R}(0)$ and
$\mathrm{var} \lambda = \langle (\lambda - \langle \lambda \rangle)^2 \rangle = {\cal R}'(z)|_{z \rightarrow 0}$, respectively.

For matrices ${\hat F}$ of the form \eref{hth}, the free random variable theory provides a number of mathematical theorems that we will exploit in this paper. In particular, one shows \cite{tulino04} that
\begin{eqnarray}
S_{{\hat F}}(z) =
\frac{1}{z + M/N} S_{\hat T} \left( \frac{N}{M} z \right),
\label{shth}
\end{eqnarray}
if ${\hat T}$ is a Hermitian nonnegative random matrix independent of ${\hat H}$ and the limits $N$, $M \rightarrow \infty$ are taken at a constant $M/N$. Using \eref{shth}, we derive a relation between the Blue function of ${\hat F}$ and the Green's function of ${\hat T}$:
\begin{eqnarray}
B_{{\hat F}}(z) &=&
\frac{1}{z}
\left\{
1 + \frac{M}{N} \left[
\frac{1}{z} {\cal G}_{{\hat T}} \left( \frac{1}{z} \right)
-1 \right] \right\}.
\label{bhth}
\end{eqnarray}

\begin{figure}
\centering{
\includegraphics[angle=-90,width=0.35\textwidth]{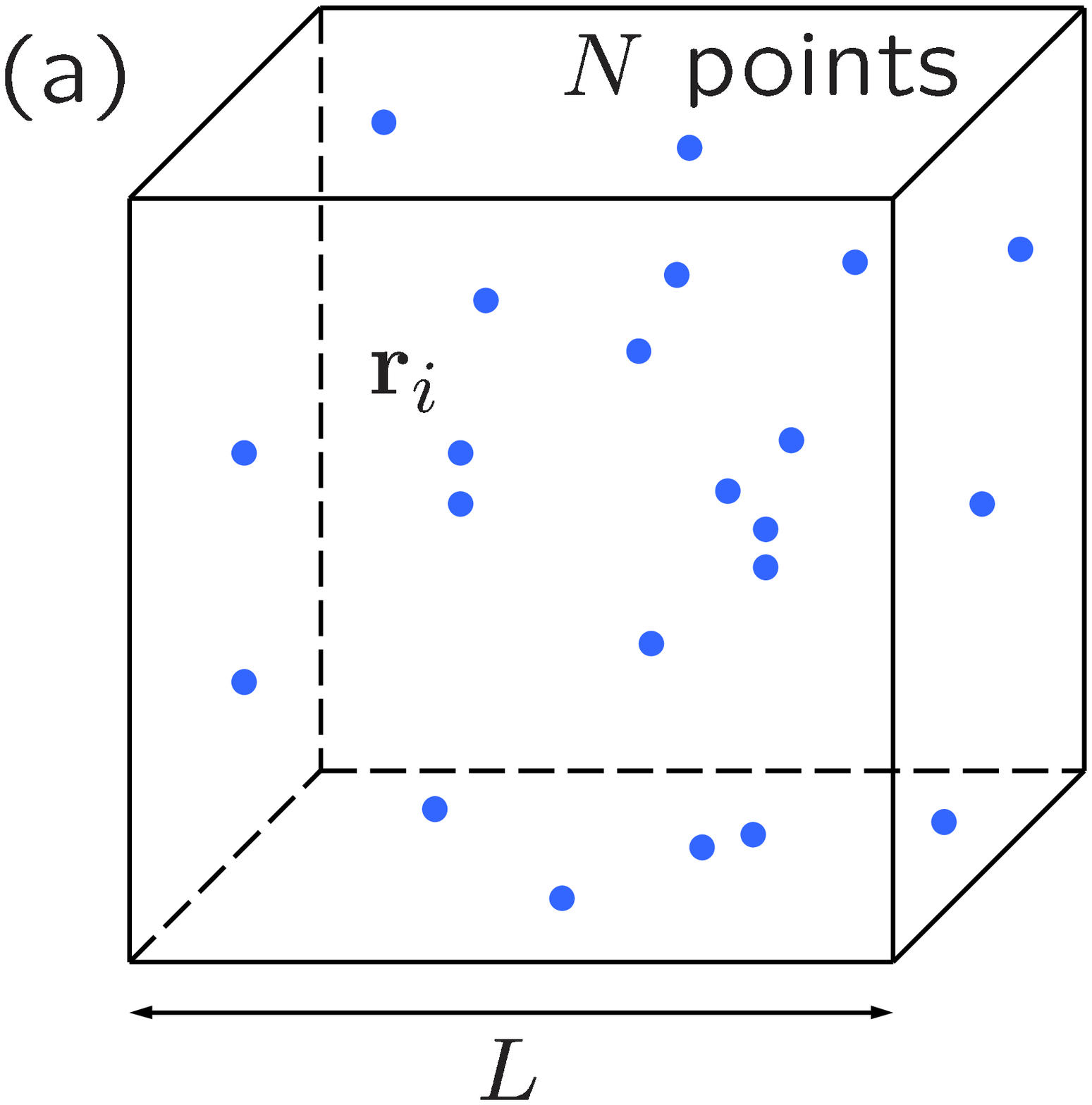}
\hspace{1.5cm}
\includegraphics[angle=-90,width=0.4\textwidth]{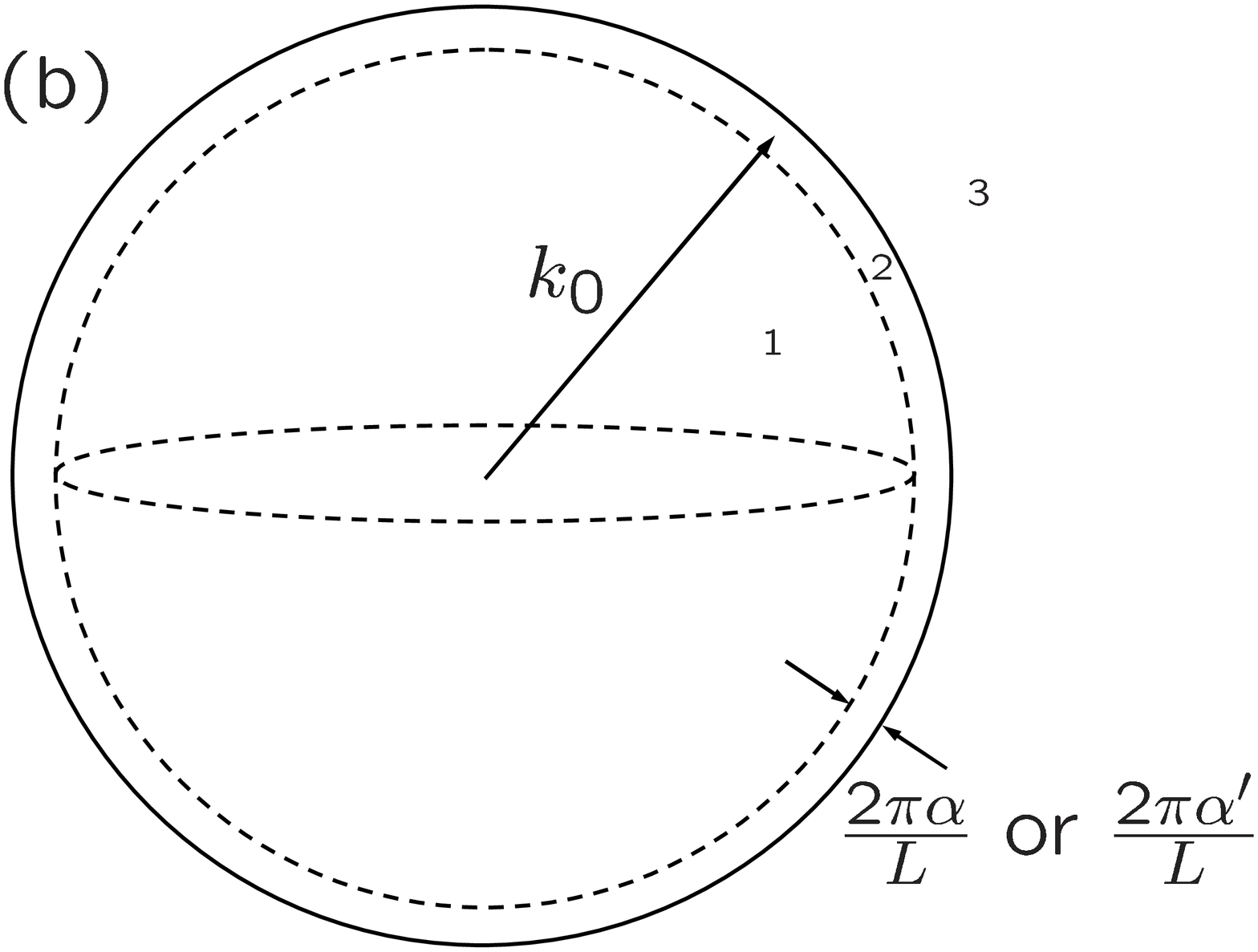}
\caption{\label{fig1}
(a) We consider $N$ points randomly distributed in a three-dimensional cube of side $L$.
(b) Regions in the Fourier space. For the sinc random matrix, only the region 2 contributes to the matrix ${\hat T}$. In contrast, for the cosc random matrix, ${\hat T}$ has contributions from the regions 1 and 3 but not from the region 2.}}
\end{figure}

A particular case that we will consider in the remainder of this paper is when the region $V$ is a square box of side $L$ [see \fref{fig1}(a)]. A convenient set of basis functions is then given by `plane waves'
\begin{eqnarray}
\psi_m(\mathbf{r}) = \frac{1}{\sqrt{V}} \rme^{\rmi \mathbf{q}_m \cdot  \mathbf{r}},
\label{planewave}
\end{eqnarray}
where $\mathbf{q}_{m} = \{q_{m_x},
q_{m_y}, q_{m_z} \}$, $q_{m_x} = m_x \Delta q$ with $m_x = \pm 1, \pm 2, \ldots$ (and similarly for $q_{m_y}$ and $q_{m_z}$), and $\Delta q = 2\pi/L$.
\Eref{tmn} is then simply a double Fourier transform of the function $f(\mathbf{r}_i, \mathbf{r}_j)$ in the box and the representation \eref{hth} stems from the Fourier series expansion of $f(\mathbf{r}_i, \mathbf{r}_j)$, without the harmonics corresponding to $\mathbf{q}_m = 0$.

\section{Eigenvalue distribution of the sinc matrix}
\label{secsinc}

We start by considering the real symmetric $N \times N$ Euclidean matrix ${\hat F} = {\hat S}$ with elements defined through the cardinal sine (sinc) function:
\begin{eqnarray}
S_{ij} = f(\mathbf{r}_i - \mathbf{r}_j) = \frac{\sin(k_0 |\mathbf{r}_i - \mathbf{r}_j|)}{
k_0 |\mathbf{r}_i - \mathbf{r}_j|}.
\label{sinc}
\end{eqnarray}
Here $k_0$ is a constant and the vectors $\mathbf{r}_i$ define positions of $N$ randomly chosen points inside a three-dimensional cube of side $L$.

The first important property of the matrix ${\hat S}$ is the positiveness of its eigenvalues: $\lambda > 0$. Indeed, the Fourier transform of the function $f(\Delta \mathbf{r})$ in \eref{sinc} is positive and hence $f(\Delta \mathbf{r})$ is a function of positive type. An Euclidean matrix defined through a function of positive type is positive definite and hence has only positive eigenvalues.
The matrix ${\hat T}$ corresponding to ${\hat S}$ can be found from \eref{tmn}:
\begin{eqnarray}
T_{mn} = \frac{N}{V^2}
\int\limits_V \rmd^3 \mathbf{r}_1
\int\limits_V \rmd^3 \mathbf{r}_2\;
\frac{\sin(k_0 |\mathbf{r}_1 - \mathbf{r}_2|)}{
k_0 |\mathbf{r}_1 - \mathbf{r}_2|}
\rme^{-\rmi \mathbf{q}_m \mathbf{r}_1
+ \rmi \mathbf{q}_n \mathbf{r}_2}.
\label{tmnsinc}
\end{eqnarray}
Unfortunately, it is impossible to calculate this double integral exactly in a box. However, introducing new variables of integration
$\mathbf{R} = \frac12 (\mathbf{r}_1 + \mathbf{r}_2)$ and
$\Delta \mathbf{r} = \mathbf{r}_2 - \mathbf{r}_1$ and limiting the integration over $\Delta \mathbf{r}$ to the region $|\Delta \mathbf{r}| < L/2\alpha$, with $\alpha \sim 1$ a numerical constant to be fixed later, we obtain an approximate result
\begin{eqnarray}
&&T_{mn} \simeq \frac{N}{V^2}
\int\limits_V \rmd^3 \mathbf{R}\;
\rme^{-\rmi (\mathbf{q}_m - \mathbf{q}_n) \mathbf{R}}
\int\limits_{|\Delta \mathbf{r}| < L/2\alpha} \rmd^3 \Delta \mathbf{r}\;
\frac{\sin(k_0 \Delta r)}{
k_0 \Delta r}
\rme^{\rmi (\mathbf{q}_m + \mathbf{q}_n) \Delta \mathbf{r}/2}
\nonumber \\
&&=
\delta_{mn} \frac{2 \pi^2 N}{k_0 q_m V} \frac{L}{2\alpha \pi} \left\{
\mathrm{sinc}\left[(q_m - k_0)\frac{L}{2\alpha} \right]
- \mathrm{sinc}\left[(q_m + k_0)\frac{L}{2\alpha} \right]
\right\}.
\label{tmnsinc2}
\end{eqnarray}
This expression is still too involved to be useful. In order to simplify it, we note that the second sinc function in \eref{tmnsinc2} is always smaller than $2 \alpha /k_0 L$ (because $q_m = |\mathbf{q}_m| > 0$ and $k_0 > 0$) and hence can be dropped in the limit of large $k_0 L \gg 1$ considered in this paper. Furthermore, because the first sinc function in \eref{tmnsinc2} is peaked around $q_m = k_0$, we replace it by a boxcar function $\Pi[(q_m - k_0) L/2 \alpha \pi]$, where $\Pi(x) = 1$ for $|x| < \frac12$ and $\Pi(x) = 0$ otherwise. The coefficient in front of $(q_m - k_0)$ in the argument of $\Pi$ is chosen to ensure that the integral of the latter over $q_m$ from 0 to $\infty$ is equal to the same integral of the sinc function. We then obtain
\begin{eqnarray}
T_{mn} &\simeq&
\frac{2 \pi^2 N}{k_0^2 V}
\frac{L}{2 \alpha \pi} \Pi\left[(q_m - k_0) \frac{L}{2 \pi \alpha} \right] \delta_{mn}
\label{tmnsinc3}
\end{eqnarray}
which is different from zero only for $\mathbf{q}_m$'s inside a spherical shell of radius $k_0$ and thickness $2\pi \alpha/L$ [i.e. in the region 2 of \fref{fig1}(b)]. In addition, for all $\mathbf{q}_m$'s inside the shell the value of $T_{mn}$ is the same and equal to $N/M$ with $M = \alpha(k_0 L)^2/\pi \gg 1$ the number of $\mathbf{q}_m$'s inside the shell. \Eref{hth} then yields
\begin{eqnarray}
{\hat S} = \frac{N}{M} {\hat H}{\hat H}^{\dagger}
\label{nmhh}
\end{eqnarray}
which is equivalent to \eref{hth} with a $M \times M$ matrix ${\hat T = (N/M) {\hat \mathbb{I}}}$,
where ${\hat \mathbb{I}}$ is the identity matrix. We then readily find
${\cal G}_{\hat T}(z) = (1/M) \Tr [z -
(N/M) {\hat \mathbb{I}}]^{-1} = (z - N/M)^{-1}$ and from \eref{bhth}: $B_{\hat S}(z) = (1-\beta z)^{-1} + 1/z$ with $\beta = N/M$. This is the Blue function of the famous Marchenko-Pastur law \cite{tulino04,marchenko67}:
\begin{eqnarray}
p(\lambda) = \left(1 - \frac{1}{\beta} \right)^+ \delta(\lambda)
+ \frac{\sqrt{(\lambda - \lambda_{\mathrm{min}})^+(\lambda_{\mathrm{max}}-
\lambda)^+}}{2\pi \beta \lambda},
\label{pmp}
\end{eqnarray}
where
$\lambda_{\mathrm{min, max}} =
(1 \mp \sqrt{\beta})^2$ and $x^+ = \mathrm{max}(x, 0)$.
The distribution of eigenvalues of the matrix \eref{sinc} is therefore parameterized by a single parameter $\beta$ equal to the variance of this distribution, as it is easy to check from \eref{pmp}: $\mathrm{var}(\lambda) = \beta$.

Although we derived \eref{pmp} using the machinery of free random variables applied to Euclidean matrices as discussed in \sref{secframe}, it represents a somewhat trivial example of application of this technique because the matrix ${\hat T}$ in \eref{hth} turns out to be proportional to the identity matrix ${\hat \mathbb{I}}$. \Eref{pmp} was first derived long before the theory of asymptotically free random variables was introduced \cite{marchenko67}. It can be established using various approaches, such as, e.g., the diagrammatic technique \cite{sengupta99}. However, to our knowledge, the fact that this distribution describes eigenvalues of the Euclidean matrix ${\hat S}$ was never noticed before. The advantage of using the free random variable theory to study Euclidean random matrices is that \eref{pmp} now appears as a special (and apparently the most trivial) case of a wide class of distributions describing matrices of the form \eref{hth}.

\begin{figure}
\vspace{2mm}
\centering{
\includegraphics[angle=-90,width=0.99\textwidth]{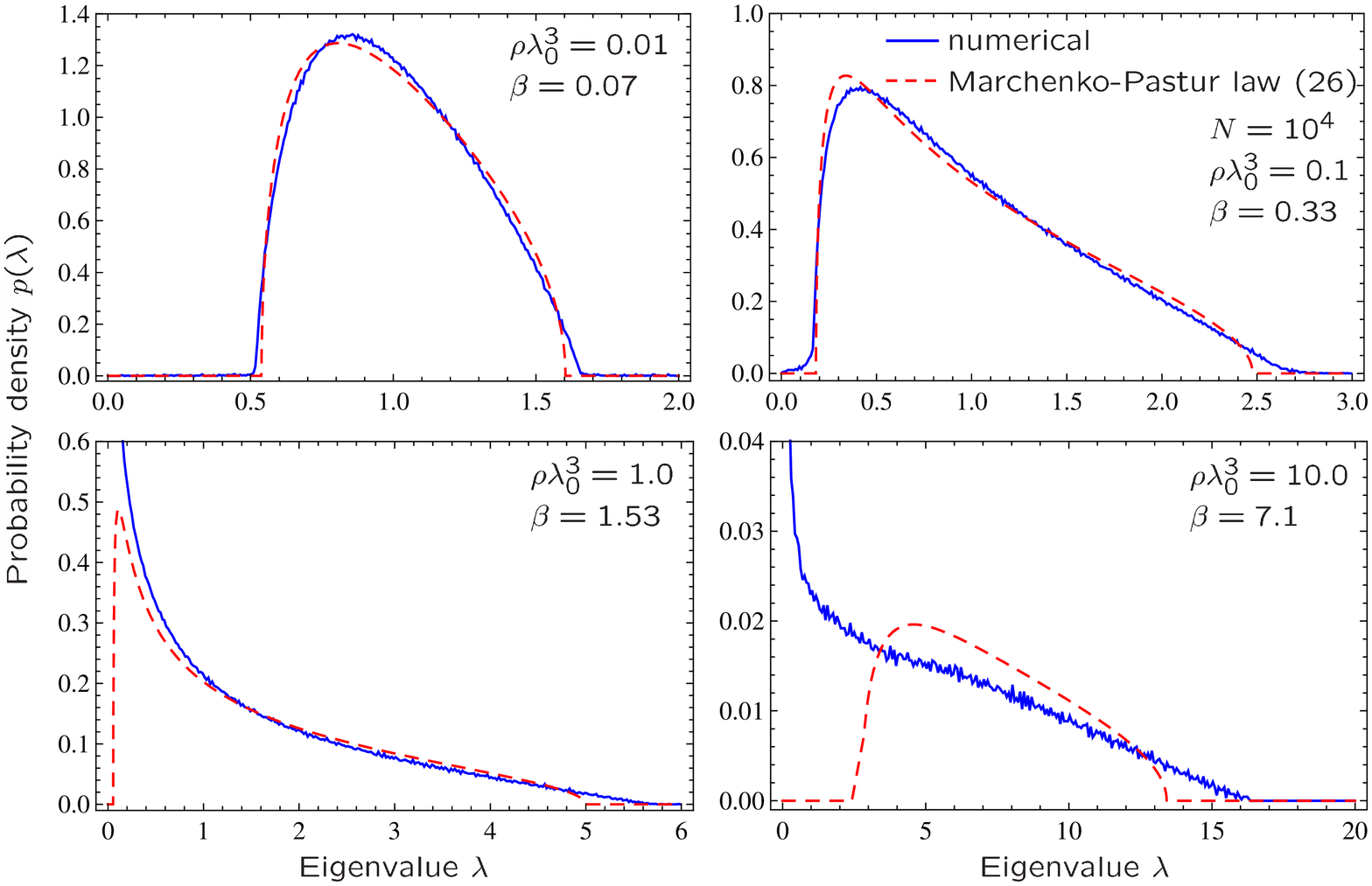}
\hspace{1.5cm}
\caption{\label{figsinc}
Probability density of eigenvalues of a square $N \times N$ Euclidean matrix ${\hat S}$ with elements
$S_{ij} = \sin(k_0 |\mathbf{r}_i - \mathbf{r}_j|)/
k_0 |\mathbf{r}_i - \mathbf{r}_j|$, where the $N$ points $\mathbf{r}_i$ are randomly chosen inside a 3D cube of side $L$. Numerical results (blue solid lines) obtained for $N = 10^4$ after averaging over 10 realizations are compared to the Marchenko-Pastur law \eref{pmp} (red dashed lines) with $\beta = 2.8 N/(k_0 L)^2$ for several densities $\rho$ of points ($\lambda_0 = 2\pi/k_0$).}}
\end{figure}

Note that despite the fact that our derivation of \eref{pmp} was based on several approximations, the average value of $\lambda$, $\langle \lambda \rangle = 1$, following from this equation is exact.
The second moment of $\lambda$ can also be found directly from \eref{sinc}. For $k_0 L \gg 1$ and in the limit $N \rightarrow \infty$ we find:
\begin{eqnarray}
\langle \lambda^2 \rangle &=& \frac{1}{N}
\langle \mathrm{Tr} {\hat S}^2 \rangle
= 1 + \frac{a N}{(k_0 L)^2},
\label{lambdasecond}
\end{eqnarray}
where the numerical constant $a$ is given by
\begin{eqnarray}
a = \frac{1}{2}
\int\limits_{\mbox{\tiny unit cube}} \rmd^3 \mathbf{u}_1
\int\limits_{\mbox{\tiny unit cube}} \rmd^3 \mathbf{u}_2
\frac{1}{|\mathbf{u}_1 - \mathbf{u}_2|^2}
&\simeq& 2.8,
\label{consta}
\end{eqnarray}
with the integrations running over the volume of a cube of unit side. By requiring that the second moment $1 + \beta$ of the distribution \eref{pmp} coincides with \eref{lambdasecond} we can now fix the value of $\alpha$ that remained arbitrary until now. We obtain $\alpha = \pi/a \simeq 1.12$ and
\begin{eqnarray}
\beta = \frac{2.8 N}{(k_0 L)^2}.
\label{beta}
\end{eqnarray}

In \fref{figsinc} we present a comparison of \eref{pmp} with the results of direct numerical simulations. The latter amount to generate $N$ random points $\mathbf{r}_i$ inside a three-dimensional cube, to use these points to define a random $N \times N$ matrix ${\hat S}$ according to \eref{sinc}, and to diagonalize ${\hat S}$ using the standard software package LAPACK \cite{lapack}. The procedure is repeated several times and a histogram of all eigenvalues $\lambda$ is created. This histogram approximates the eigenvalue distribution $p(\lambda)$. As we see from \fref{figsinc}, the agreement between numerical results and the Marchenko-Pastur law \eref{pmp}  is good for $\beta < 1$ but \eref{pmp} fails to describe $p(\lambda)$ when $\beta$ becomes larger than unity. The reason for this is easy to understand if we go back to \eref{tmnsinc}, \eref{tmnsinc2} and \eref{tmnsinc3}. Indeed, when we approximate the result of integration in \eref{tmnsinc} by \eref{tmnsinc3}, we reduce the infinite-size matrix ${\hat T}$ to a matrix of finite size $M \times M$. By definition, the rank of the latter matrix is inferior or equal to $M$. The rank of ${\hat S} = {\hat H} {\hat T} {\hat H}^{\dagger}$ cannot be larger than the rank of ${\hat T}$ and hence is also bounded by $M$ from above when we use \eref{tmnsinc3}. When $\beta > 1$, implying $M < N$, the representation \eref{hth} only gives us access to $M$ of $N$ eigenvalues of ${\hat S}$, which is not sufficient to reconstruct the probability density $p(\lambda)$. In order to access the regime of $\beta > 1$ one needs to find a better approximation to \eref{tmnsinc} than \eref{tmnsinc3}.

Note that the eigenvalue distribution of the matrix ${\hat S}$ has been studied numerically by Akkermans {\it et al.} in the context of light propagation in atomic gases (see figure 1 of \cite{akker08}) without proposing any analytical approximation to it. The parameter $\beta \sim N/(k_0 L)^2$ has been introduced in that work as a ratio of the number of atoms $N$ to the number of transverse optical modes $N_{\perp} \propto (k_0 L)^2$. The same parameter appeared in \cite{svid08,scully09,svid09,svid10} as a superradiant decay rate in a cold atomic gas.
Hence the results of this section complement and extend the works \cite{svid08,scully09,svid09,svid10,akker08}.

\section{Eigenvalue distribution of cosc matrix}
\label{seccosc}

Let us now consider an Euclidean random matrix with elements defined using the cardinal cosine (cosc) function:
\begin{eqnarray}
C_{ij} = f(\mathbf{r}_i - \mathbf{r}_j) =
(1 - \delta_{ij})
\frac{\cos(k_0 |\mathbf{r}_i - \mathbf{r}_j|)}{
k_0 |\mathbf{r}_i - \mathbf{r}_j|}.
\label{cosc}
\end{eqnarray}
The prefactor $1-\delta_{ij}$ allows us to deal with the divergence of the function $\cos(x)/x$ for $x \rightarrow 0$. However, in the beginning of our analysis we will ignore this prefactor and will use
$C_{ij} = \cos(k_0 |\mathbf{r}_i - \mathbf{r}_j|)/
k_0 |\mathbf{r}_i - \mathbf{r}_j|$ for all $i,j$ (the diagonal elements of ${\hat C}$ are thus infinite). Proceeding as in the previous section, we find
\begin{eqnarray}
T_{mn} \simeq
\frac{4 \pi N}{k_0 V} \frac{1}{q_m^2 - k_0^2} \delta_{mn}
\label{tmncosc}
\end{eqnarray}
under the same approximations as in \eref{tmnsinc2} (i.e., we extended integration over $\Delta \mathbf{r}$ to the whole space). The matrix ${\hat T}$ defined by \eref{tmncosc} has infinite size.

The divergence of \eref{tmncosc} for $q_m \rightarrow k_0$ can be traced back to the neglect of the finiteness of the volume $V$ when extending integration over $\Delta \mathbf{r}$ to the whole space. Taking into account the fact that $\Delta r$ cannot exceed a maximum value of the order of $L$, the divergence is regularized and the resulting $T_{mn}$ changes sign rapidly but continuously in a strip of width $\sim 1/L$ around $q_m = k_0$. In the following, we will neglect the contribution of $\mathbf{q}_m$'s inside the spherical shell corresponding to this strip because (i) the shell has small thickness in the limit of $k_0 L \gg 1$ that we are interested in and (ii) $\mathbf{q}_m$'s situated symmetrically with respect to the surface $q_m = k_0$ yield contributions of roughly equal magnitudes but opposite signs which approximately cancel. More precisely, we will exclude a shell of thickness $2 \pi \alpha'/L$ around $q_m = k_0$ and will use \eref{tmncosc} outside this shell [see \fref{fig1}(b)]. The numerical constant $\alpha' \sim 1$ will be fixed later.
The matrix ${\hat C}$ therefore takes the form:
\begin{eqnarray}
{\hat C} = -{\hat H}^{(1)} {\hat T}^{(1)} {\hat H}^{(1)\dagger}
+ {\hat H}^{(3)} {\hat T}^{(3)} {\hat H}^{(3)\dagger}.
\label{hth3}
\end{eqnarray}
Here the first term describes the contribution of
$q_m < k_0 - \alpha' \pi/L$ with
$T^{(1)}_{mn} = 4 \pi N/[k_0 V (k_0^2 - q_m^2)] \delta_{mn}$.
The matrix ${\hat T}^{(1)}$ is a diagonal square matrix of size $M_1 \simeq (4\pi/3) (k_0 - \alpha' \pi/L)^3
(L/2\pi)^3$ obtained by dividing the volume of a sphere of radius $k_0 - \alpha' \pi/L$ [region 1 corresponding to the inner sphere in \fref{fig1}(b)] by the volume $(2\pi/L)^3$ associated with a single mode. The second term in \eref{hth3} corresponds to $q_m > k_0 + \alpha' \pi/L$ [region 3 in \fref{fig1}(b)]. The matrix ${\hat T}^{(3)}$ is, again, diagonal, with elements $T^{(3)}_{mn} = 4 \pi N/[k_0 V (q_m^2 - k_0^2)] \delta_{mn}$  but, in contrast to ${\hat T}^{(1)}$, has infinite size. We will treat this matrix as a finite-size matrix of size $M_3 \simeq (4\pi/3) [q_{\mathrm{max}}^3 - (k_0 + \alpha' \pi/L)^3 ] (L/2\pi)^3$, corresponding to taking into account only $q_m \leq q_{\mathrm{max}}$. The limit of $q_{\mathrm{max}} \rightarrow \infty$ will be taken at the end. The minus sign in front of the first term in \eref{hth3} was introduced to work with a positive-definite matrix ${\hat T}^{(1)}$.

The Green's function of the matrix ${\hat T}^{(1)}$ is
\begin{eqnarray}
{\cal G}_{{\hat T}^{(1)}}(z) &=& \frac{1}{M_1} \mathrm{Tr} \frac{1}{z - {\hat T}^{(1)}}
\nonumber \\
&=& \frac{1}{M_1} \sum\limits_{q_m < k_0 - \alpha' \pi/L}
\frac{1}{z - \frac{4 \pi N}{k_0 V} \frac{1}{k_0^2 - q_m^2}}
\nonumber \\
&\simeq&
\frac{4 \pi N}{M_1 (\rho \lambda_0^3)}
\int\limits_{0}^{1 - \frac{\alpha' \pi}{k_0 L}}
\rmd \kappa\; \kappa^2 \frac{1}{z - \frac{\rho \lambda_0^3}{2 \pi^2} \frac{1}{1- \kappa^2}},
\label{gt1}
\end{eqnarray}
where the wavelength $\lambda_0 = 2\pi/k_0$ and $\rho \lambda_0^3$ is the average number of points $\mathbf{r}_i$ per wavelength cube.
The last line of this equation was obtained in the limit of $k_0 L \gg 1$ by approximately replacing the summation over a set of discrete wavevectors $\mathbf{q}_m$ by an integration over $\boldsymbol{\kappa} = \mathbf{q}_m/k_0$. The integral in \eref{gt1} can be evaluated yielding
\begin{eqnarray}
{\cal G}_{{\hat T}^{(1)}}(z) &=&
\frac{2 N}{\pi M_1 z}
\left\{
\left( 1 - \frac{\alpha' \pi}{k_0 L} \right)
\left[ \frac{2\pi^2}{3\rho\lambda_0^3}
\left( 1 - \frac{\alpha' \pi}{k_0 L} \right)^2 -
\frac{1}{z} \right] \right.
\nonumber \\
&+& \left.
\frac{1}{z}
\sqrt{\frac{\rho \lambda_0^3}{2\pi^2 z} - 1}\;
\arctan \frac{ 1 - \frac{\alpha' \pi}{k_0 L}}{\sqrt{\frac{\rho \lambda_0^3}{2\pi^2 z} - 1}} \right\}.
\label{gt1a}
\end{eqnarray}

A similar calculation can be performed for the Green's function of ${\hat T}^{(3)}$ except that the integration in \eref{gt1} extends from $1 + \alpha' \pi/k_0 L$ to $\kappa_{\mathrm{max}}$, $M_1$ is replaced by $M_3$, and $1 - \kappa^2$ in the integrand of \eref{gt1} --- by $\kappa^2 - 1$. We find
\begin{eqnarray}
{\cal G}_{{\hat T}^{(3)}}(z) &=&
\frac{2 N}{\pi M_3 z}
\left\{
\left[ \kappa_{\mathrm{max}} - \left(1 + \frac{\alpha' \pi}{k_0 L} \right) \right]
\left[ \frac{2\pi^2}{3\rho\lambda_0^3}
\left( \kappa_{\mathrm{max}}^2
\right. \right. \right.
\nonumber \\
&+& \left. \left. \left.
\left( 1 + \frac{\alpha' \pi}{k_0 L} \right)
\left( \kappa_{\mathrm{max}} + 1 + \frac{\alpha' \pi}{k_0 L}
\right) \right)
+ \frac{1}{z} \right] \right.
\nonumber \\
&+& \left.
\frac{1}{z}
\sqrt{-\frac{\rho \lambda_0^3}{2\pi^2 z} - 1}
\left[
\arctan \frac{ 1 + \frac{\alpha' \pi}{k_0 L}}{\sqrt{-\frac{\rho \lambda_0^3}{2\pi^2 z} - 1}}
\right. \right. \nonumber \\
&-& \left. \left.
\arctan \frac{\kappa_{\mathrm{max}}}{\sqrt{-\frac{\rho \lambda_0^3}{2\pi^2 z} - 1}} \right] \right\}.
\label{gt3}
\end{eqnarray}

Because the two terms in \eref{hth3} correspond to contributions of different parts of $\mathbf{q}$-space, they are asymptotically free and hence the Blue function of their sum (i.e. of the matrix ${\hat C}$) can be found as a sum of their respective Blue functions.
The Blue functions of ${\hat H}^{(1)} {\hat T}^{(1)} {\hat H}^{(1)\dagger}$ and ${\hat H}^{(3)} {\hat T}^{(3)} {\hat H}^{(3)\dagger}$ are found using \eref{bhth}, whereas
the Blue function of
$-{\hat H}^{(1)} {\hat T}^{(1)} {\hat H}^{(1)\dagger}$ is equal to $-B_{{\hat H}^{(1)} {\hat T}^{(1)} {\hat H}^{(1)\dagger}}(-z)$. The Blue function of the sum
${\hat C} = -{\hat H}^{(1)} {\hat T}^{(1)} {\hat H}^{(1)\dagger} +
{\hat H}^{(3)} {\hat T}^{(3)} {\hat H}^{(3)\dagger}$ is a sum of individual Blue functions, minus $1/z$:
\begin{eqnarray}
B_{\hat C}(z) &=& -B_{{\hat H}^{(1)} {\hat T}^{(1)} {\hat H}^{(1)\dagger}}(-z) + B_{{\hat H}^{(3)} {\hat T}^{(3)} {\hat H}^{(3)\dagger}}(z) - 1/z
\nonumber \\
&=& \frac{1}{z} + \frac{2 \kappa_{\mathrm{max}}}{\pi} - \frac{\rho \lambda_0^3}{2 \pi^2 \beta'} + \frac{2}{\pi}
\sqrt{-1 - \frac{\rho \lambda_0^3}{2\pi^2} z}
\nonumber \\
&\times& \left[
\arctan \frac{1 + \frac{\rho \lambda_0^3}{8 \pi \beta'}}{\sqrt{-1 - \frac{\rho \lambda_0^3}{2\pi^2} z}} -
\arctan \frac{1 - \frac{\rho \lambda_0^3}{8 \pi \beta'}}{\sqrt{-1 - \frac{\rho \lambda_0^3 }{2\pi^2} z}}
\right.
\nonumber \\
&-& \left. \arctan \frac{\kappa_{\mathrm{max}}}{\sqrt{-1 - \frac{\rho \lambda_0^3}{2\pi^2} z}} \right].
\label{bluefull}
\end{eqnarray}
where $\beta' = \pi N/\alpha' (k_0 L)^2$.

The final step consists in taking the limit $\kappa_{\mathrm{max}} \rightarrow \infty$. We now recall that up to now we ignored the fact that the matrix ${\hat C}$ had zero diagonal elements $C_{ii} = 0$. Instead, we considered a matrix with infinitely large diagonal elements. Such a matrix naturally has infinite eigenvalues and to go back to the case of $C_{ii} = 0$ we have to shift the eigenvalues to the left. To determine the exact shift, we compute the average of $\lambda$ from \eref{bluefull} using \eref{momentb} and subtract it from \eref{bluefull} because we know that for the matrix ${\hat C}$ defined by \eref{cosc}, $\langle \lambda \rangle = (1/N) \langle \Tr {\hat C} \rangle = 0$ exactly \footnote{Subtracting a constant from the Blue function $B(z)$ results in shifting the eigenvalue distribution $p(\lambda)$.}. We then compute the second moment $\langle \lambda^2 \rangle$ and require that its value in the limit of $\rho \lambda_0^3/\beta' \propto 1/k_0 L \rightarrow 0$ is equal to $\beta$ defined by \eref{beta}. This fixes $\beta' = (\pi^2/4) \beta$ corresponding to $\alpha' \simeq 0.45$. The final expression for the Blue function of ${\hat C}$ is
\begin{eqnarray}
B_{\hat C}(z)
&=& \frac{1}{z} - \frac{2}{\pi} \mathrm{arccoth} \frac{4 \pi^3 \beta}{\rho \lambda_0^3} + \frac{2}{\pi}
\sqrt{-1 - \frac{\rho \lambda_0^3}{2\pi^2} z}
\nonumber \\
&\times& \left[
\arctan \frac{1 + \frac{\rho \lambda_0^3}{2 \pi^3 \beta}}{\sqrt{-1 - \frac{\rho \lambda_0^3}{2\pi^2} z}} -
\arctan \frac{1 - \frac{\rho \lambda_0^3}{2 \pi^3 \beta}}{\sqrt{-1 - \frac{\rho \lambda_0^3 }{2\pi^2} z}}
- \frac{\pi}{2} \right].
\label{bluefull1}
\end{eqnarray}
The Green's function ${\cal G}_{\hat C}(z)$ can be found from this equation by solving $B_{\hat C}[{\cal G}(z)] = z$ which, for the general case, we do numerically.

\begin{figure}
\vspace{3mm}
\centering{
\includegraphics[angle=-90,width=0.99\textwidth]{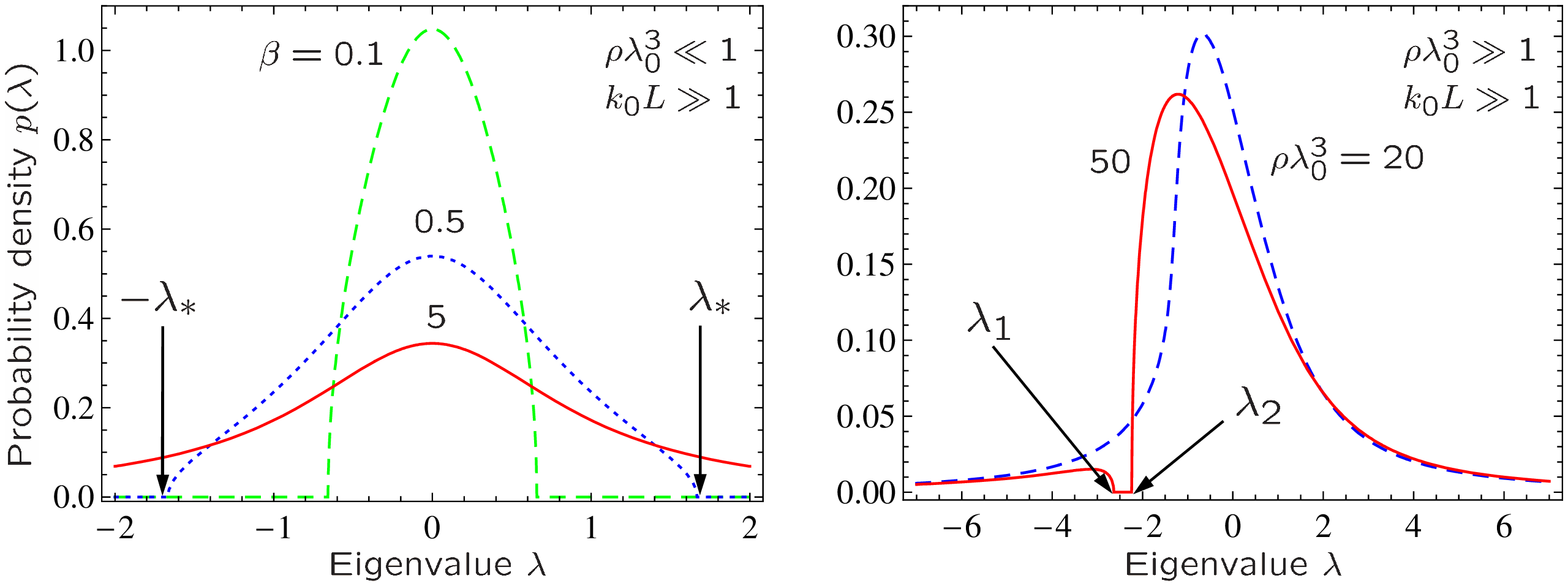}
\hspace{1.5cm}
\caption{\label{figcosclimit}
Probability density of eigenvalues of a square $N \times N$ Euclidean matrix ${\hat C}$ with elements
$C_{ij} = (1-\delta_{ij}) \cos(k_0 |\mathbf{r}_i - \mathbf{r}_j|)/
k_0 |\mathbf{r}_i - \mathbf{r}_j|$, where the $N$ points $\mathbf{r}_i$ are randomly chosen inside a 3D cube of side $L$. The left panel corresponds to the low-density limit and is obtained using \eref{bluerho0} with $\beta = 0.1$, 0.5 and 5. The distributions are symmetric and vanish for $|\lambda| > \lambda_*$ with $\lambda_*$ given by \eref{lambdastar}.
The right panel illustrates our equation \eref{bluerholarge} obtained in the high-density limit for two densities $\rho \lambda_0^3 = 20$ and 50. For $\rho \lambda_0^3 > 30.3905$ the distribution develops a gap in between $\lambda_1$ and $\lambda_2$ given by \eref{lambda1} and \eref{lambda2}, respectively.
}}
\end{figure}

Let us consider the low-density limit of \eref{bluefull1}, $\rho \lambda_0^3 \ll 1$. For large box size $L \gg 1/k_0$ the arguments of $\arctan$ functions in \eref{bluefull1} are close to $-\rmi$. They can be thus expanded in series in the vicinity of this point. In the resulting expression we take  the limits of $\rho \lambda_0^3 \rightarrow 0$ and $\rho \lambda_0^3/\beta \sim
1/k_0 L \rightarrow 0$ to obtain
\begin{eqnarray}
B_{\hat C}(z) = \frac{1}{z} - \frac{1}{\pi} \ln \frac{1 - {\frac{\pi}{2} \beta z}}{1 + {\frac{\pi}{2} \beta z}},
\;\;\; \rho \lambda_0^3 \ll 1.
\label{bluerho0}
\end{eqnarray}
This expression has two important limits. For $\beta \ll 1$ we find $B_{\hat C}(z) = \beta z + 1/z$ which is the Blue function of the Wigner semi-circle law
$p(\lambda) = \sqrt{4 \beta - \lambda^2}/2 \pi \beta$.
In the opposite limit of $\beta \gg 1$ we have $B_{\hat C}(z) = -\rmi + 1/z$, which corresponds to the Cauchy distribution $p(\lambda) = 1/[\pi (1 + \lambda^2)]$. \Eref{bluerho0} therefore describes a transition from the Wigner semi-circle law at $\beta \ll 1$ to the Cauchy distribution at $\beta \rightarrow \infty$.
The eigenvalue distribution following from \eref{bluerho0} is always symmetric with respect to $\lambda = 0$ and vanishes for $|\lambda| > \lambda_*$ (see the left panel of \fref{figcosclimit}). The latter can be found by using the relation $p(\lambda) \propto \mathrm{Im} {\cal G}(z = \lambda + \rmi \epsilon)$ and the link between ${\cal G}(z)$ and $B(z)$.
Simple reasoning shows that the boundary $\lambda_*$ of the domain of existence of eigenvalues is the solution of equation $B_{\hat C}'(z) = 0$ \cite{zee96}:
\begin{eqnarray}
\lambda_* = \sqrt{\beta \left( 1+\frac{\pi^2}{4} \beta \right)} +
\frac{2}{\pi} \mathrm{arccoth} \sqrt{1 + \frac{4}{\pi^2 \beta}}.
\label{lambdastar}
\end{eqnarray}
This equation simplifies to $\lambda_* = 2\sqrt{\beta}$ for $\beta \ll 1$ and to $\lambda_* = \frac{\pi}{2} \beta$ for $\beta \gg 1$.

Another important limit of \eref{bluefull1} is that of high density $\rho \lambda_0^3 \gg 1$ of points in a large box $L \gg 1/k_0$. In this limit the arguments of $\arctan$ functions in \eref{bluefull1} are small and we can put $\arctan x \simeq x$. Taking the limit of $\rho \lambda_0^3/\beta' \sim 1/k_0 L \rightarrow 0$ we then obtain
\begin{eqnarray}
B_{\hat C}(z) = \frac{1}{z} + \rmi \sqrt{1 + \frac{\rho \lambda_0^3}{2 \pi^2} z},
\;\;\; \rho \lambda_0^3 \gg 1.
\label{bluerholarge}
\end{eqnarray}
For $\rho \lambda_0^3$ below a critical value $(\rho \lambda_0^3)_c = 30.3905$ the eigenvalue distribution corresponding to \eref{bluerholarge} is asymmetric but bell-shaped, similarly to the case of low density. For $\rho \lambda_0^3 > (\rho \lambda_0^3)_c$, however, the distribution develops a gap: $p(\lambda) = 0$ for $\lambda_1 < \lambda < \lambda_2$, where $\lambda_{1,2} = B_{\hat C}(z_{1,2})$ with $z_{1,2}$ being solutions of $B_{\hat C}'(z) = 0$ (see the right panel of \fref{figcosclimit}). In the limit of $\rho \lambda_0^3 \gg (\rho \lambda_0^3)_c$ we have
\begin{eqnarray}
\lambda_1 &\simeq& -\frac{\rho \lambda_0^3}{2 \pi^2} -
\frac{\pi^2}{2 \rho \lambda_0^3},
\label{lambda1}
\\
\lambda_2 &\simeq& -\frac{3}{2 \pi^{2/3}} (\rho \lambda_0^3)^{1/3} +
\frac{\pi^{2/3}}{2 (\rho \lambda_0^3)^{1/3}} +
\frac{\pi^2}{6 \rho \lambda_0^3}.
\label{lambda2}
\end{eqnarray}

\begin{figure}
\vspace{2mm}
\centering{
\includegraphics[angle=-90,width=0.99\textwidth]{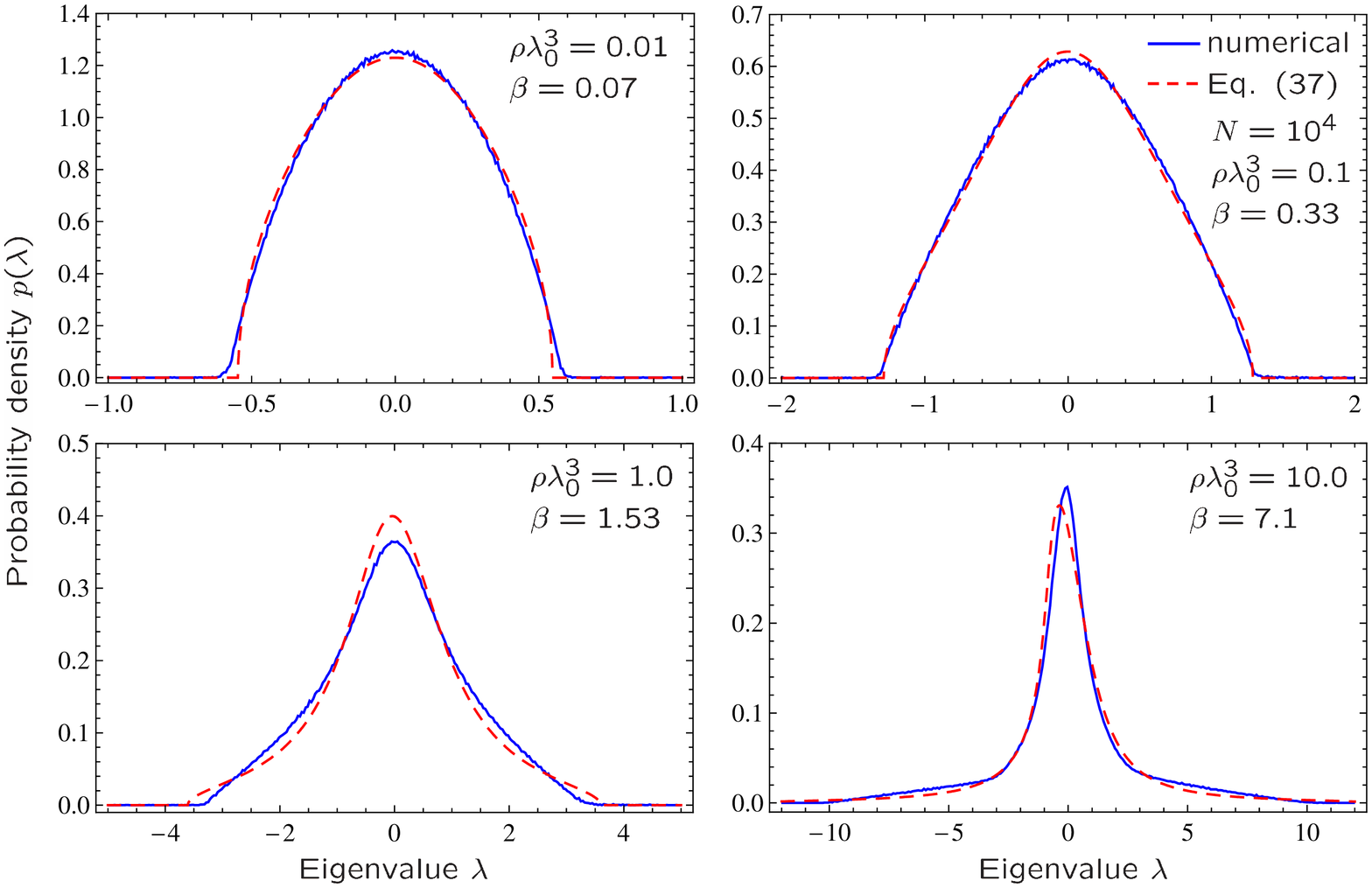}
\hspace{1.5cm}
\caption{\label{figcosc}
Probability density of eigenvalues of a square $N \times N$ Euclidean matrix ${\hat C}$ with elements
$C_{ij} = (1-\delta_{ij}) \cos(k_0 |\mathbf{r}_i - \mathbf{r}_j|)/
k_0 |\mathbf{r}_i - \mathbf{r}_j|$, where the $N$ points $\mathbf{r}_i$ are randomly chosen inside a 3D cube of side $L$. Numerical results (blue solid lines) obtained for $N = 10^4$ after averaging over 10 realizations are compared to our equation \eref{bluefull1} (red dashed lines) with $\beta = 2.8 N/(k_0 L)^2$ for several densities $\rho$ of points ($\lambda_0 = 2\pi/k_0$).}}
\end{figure}

In \fref{figcosc} we compare $p(\lambda)$ following from \eref{bluefull1} with the results of numerical simulations.  We find the Green's function ${\cal G}_{\hat C}(z)$ by solving the equation $B_{\hat C}[{\cal G}_{\hat C}(z)] = z$ numerically and then evaluate the probability distribution of eigenvalues $p(\lambda)$ with the help of \eref{pimg}. When $\beta \rightarrow 0$, the distribution $p(\lambda)$ tends to the Wigner semi-circle law. In contrast, for large $\beta > 1$ it resembles a Cauchy distribution. A good agreement between numerical results and \eref{bluefull1} is observed not only for $\beta < 1$ (similarly to the case of sinc matrix in \sref{secsinc}) but for $\beta > 1$ as well. Note that in contrast to the Marchenko-Pastur law \eref{pmp} parameterized by a single parameter $\beta$, the Green's function \eref{bluefull1} and the corresponding probability distribution depend on two parameters $\beta$ and $\rho \lambda_0^3$. A good agreement between \eref{bluefull1} and numerical simulations is obtained at low densities $\rho \lambda_0^3 < 30$ (see \fref{figcosc}). In contrast, at higher densities $\rho \lambda_0^3 \gtrsim 30$ (not shown) the probability distribution following from \eref{bluefull1} develops a gap that is not present in numerical results. Interestingly, this gap in the probability distribution appears at the same density $\rho \lambda_0^3 \approx 30$ for all $\beta$.

\section{Eigenvalue distribution of cosc + $\rmi$ sinc matrix}
\label{secfree}

The matrices ${\hat C}$ and ${\hat S}$ can be combined in a single complex non-Hermitian matrix: ${\hat C} + \rmi ({\hat S} - {\hat \mathbb{I}})$. The theory of free random variables \cite{voiculescu92} allows one to study the statistical distribution of the complex eigenvalues of this matrix based on the properties of the matrices ${\hat C}$ and ${\hat S}$ that we considered in the previous sections \cite{jarosz06}. This, however, requires asymptotic freeness of ${\hat C}$ and ${\hat S}$. Unfortunately, the matrices ${\hat S}$ and ${\hat C}$ defined by \eref{sinc} and \eref{cosc} through {\it the same\/} set of points $\{ \mathbf{r}_i \}$ turn out to be not asymptotically free. We therefore start our study of non-Hermitian Euclidean random matrices by the case of a matrix ${\hat X } = {\hat C} + \rmi ({\hat S}' - {\hat \mathbb{I}})$, where {\it two  different and independent\/} sets of points $\{ \mathbf{r}_i \}$ and $\{ \mathbf{r}_i' \}$ are used to define the real and imaginary parts of ${\hat X}$:
\begin{eqnarray}
C_{ij} &=& (1 - \delta_{ij})
\frac{\cos(k_0 |\mathbf{r}_i - \mathbf{r}_j|)}{
k_0 |\mathbf{r}_i - \mathbf{r}_j|},
\nonumber \\
S_{ij}' &=&
\frac{\sin(k_0 |\mathbf{r}_i' - \mathbf{r}_j'|)}{
k_0 |\mathbf{r}_i' - \mathbf{r}_j'|}.
\label{xc}
\end{eqnarray}
The matrix ${\hat X}$ defined in this way is similar to the matrix ${\hat G}$ defined in the introduction except that it has no correlation between its real and imaginary parts.
Using the definition of asymptotic freeness \cite{voiculescu92,tulino04} it is easy to check that the matrices ${\hat C}$ and ${\hat S}'$ are asymptotically free, in agreement with the intuitive definition of freeness as statistical independence.
One can easily show that for the same reason as the one that ensured positiveness of the eigenvalues of the matrix ${\hat S}$ in \sref{secsinc}, the complex eigenvalues $\lambda$ of the matrix ${\hat X}$ obey $\mathrm{Im} \lambda > -1$.

For non-Hermitian matrices, the Green's function loses its analyticity inside two-dimensional domains (`islands') on the complex plane, instead of segments of the real axis in the Hermitian case. In \cite{jarosz06} Jarosz and Nowak provide a simple algorithm, based on the algebra of quaternions, to calculate the non-holomorphic Green's function ${\cal G}_{\hat X}(z)$ and the correlator of left $|L_i \rangle$ and right $|R_i \rangle$ eigenvectors \cite{chalker98} ${\cal C}_{\hat X}(z)=-(\pi/N) \langle
\sum_{i=1}^N \langle L_i|L_i \rangle
\langle R_i|R_i \rangle \delta(z-\lambda_i) \rangle$ inside these domains for any non-Hermitian matrix of the form $\hat X = \hat H_1 + \rmi \hat H_2$, where $\hat H_1$ and $\hat H_2$ are two asymptotically free Hermitian matrices with known Blue functions. In our case, ${\hat H}_1 = {\hat C}$ and ${\hat H}_2 = {\hat S}' - {\hat \mathbb{I}}$. In the limit of $\beta \ll 1$, the Blue functions are $B_1(z) = \beta z + 1/z$ (\sref{seccosc}) and $B_2(z) = 1/(1 - \beta z) - 1 + 1/z$ (\sref{secsinc}). ${\cal G}_{\hat X}(z)$ and ${\cal C}_{\hat X}(z)$ can be then found analytically:
\begin{eqnarray}
{\cal G}_{\hat X}(z=x+\rmi y) &=& \frac{x}{2\beta}-
\frac{\rmi}{2}\left[\frac{y}{\beta(1+y)}+\frac{1}{2+y}\right],
\label{greenx}
 \\
{\cal C}_{\hat X}(z=x+\rmi y) &=& \left(\frac{x}{2\beta}\right)^2 + \frac{1}{4}\left[ \frac{y}{\beta(1+y)} - \frac{1}{2 + y} \right]^2
\nonumber \\
&-& \frac{1}{\beta(1 + y)(2 + y)}.
\label{corrx}
\end{eqnarray}
The correlator \eref{corrx} must vanish on the borderline of the eigenvalue domains. We therefore readily obtain an equation for the borderline of the domain of existence of eigenvalues of ${\hat X}$ on the complex plane:
\begin{eqnarray}
x^2 + \left( \frac{y}{1 + y} - \frac{\beta}{2 + y} \right)^2 - \frac{4 \beta}{(1 + y)(2 + y)} = 0,
\label{domain}
\end{eqnarray}
where $x = \mathrm{Re} \lambda$ and $y = \mathrm{Im} \lambda$. The probability density inside this domain is
\begin{eqnarray}
p(x,y) &=&
\frac{1}{2\pi}\left[\partial_x\mathrm{Re}\,{\cal G}_{\hat X}(x,y)-\partial_y\mathrm{Im}\,{\cal G}_{\hat X}(x,y)\right]
\nonumber \\
&=& \frac{1}{4 \pi} \left[
\frac{1}{\beta} + \frac{1}{\beta (1+y)^2} - \frac{1}{(2+y)^2} \right].
\label{pfree}
\end{eqnarray}

\begin{figure}
\vspace{2mm}
\centering{
\includegraphics[angle=-90,width=0.95\textwidth]{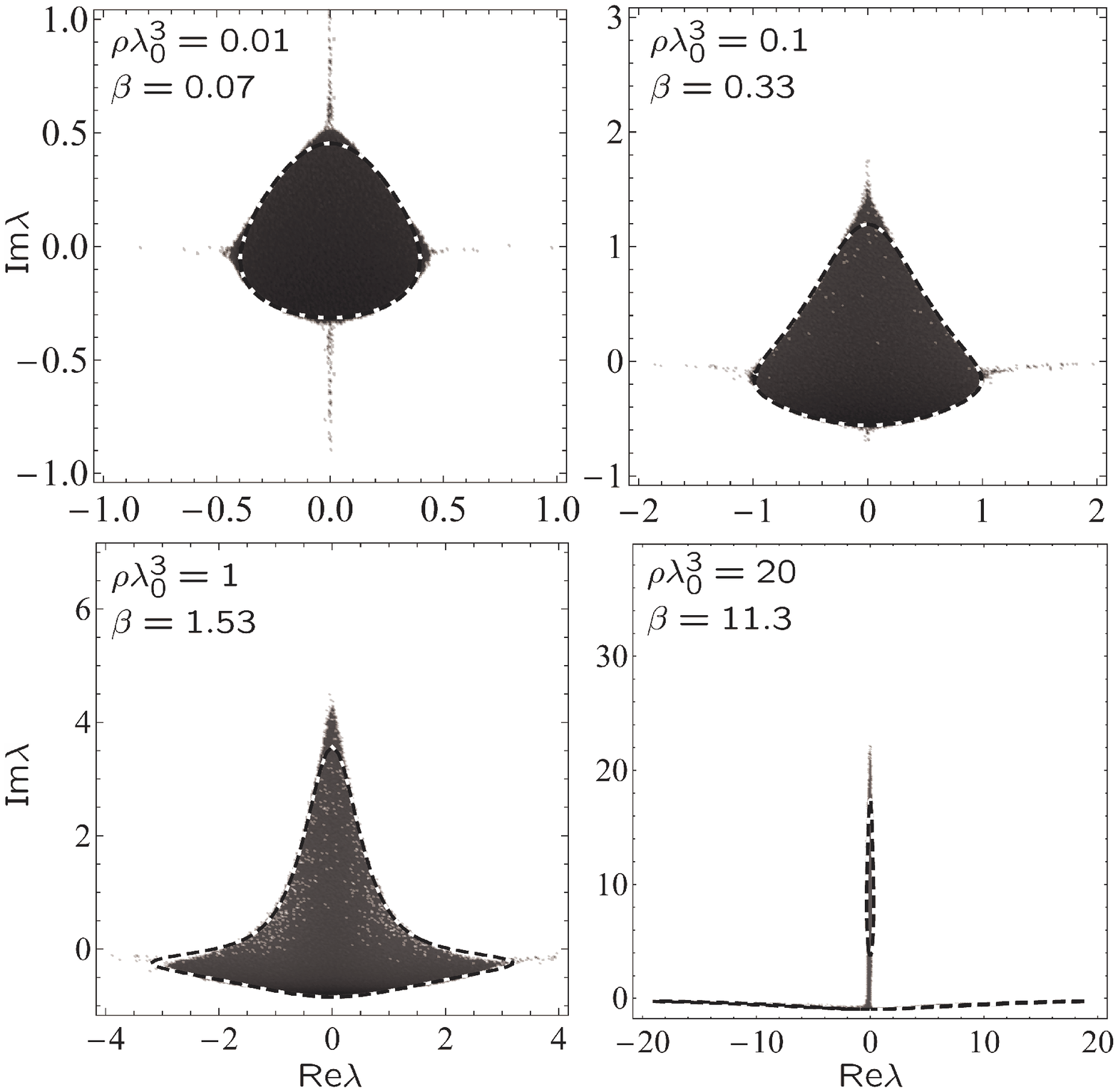}
\hspace{1.5cm}
\caption{\label{figfree}
Density plot of the logarithm of the probability density of eigenvalues $\lambda$ of a square $N \times N$ Euclidean matrix ${\hat X}$ with elements
$X_{ij} = (1-\delta_{ij}) [\cos(k_0 |\mathbf{r}_i - \mathbf{r}_j|)/
k_0 |\mathbf{r}_i - \mathbf{r}_j|
+ \rmi \sin(k_0 |\mathbf{r}_i' - \mathbf{r}_j'|)/
k_0 |\mathbf{r}_i' - \mathbf{r}_j'| ]$
at 4 different densities $\rho$ of points ${\mathbf{r}_i}$, ${\mathbf{r}_i}'$ per wavelength $\lambda_0 = 2\pi/k_0$ cube. $2N = 2 \times 10^4$ points $\mathbf{r}_i$ and $\mathbf{r}_i'$ ($i = 1, \ldots, N$) are randomly chosen inside a 3D cube; the probability distributions are estimated from 10 realizations of $\{\mathbf{r}_i \}$ and $\{\mathbf{r}_i' \}$. Dashed lines show the domain of existence of eigenvalues following from the free probability theory.}}
\end{figure}

A better model for the Blue function of the matrix ${\hat C}$ is \eref{bluerho0}. If we use this equation instead of $B_1(z) = \beta z + 1/z$ above, analytic calculation becomes impossible but we can still compute ${\cal G}_{\hat X}(z)$ and ${\cal C}_{\hat X}(z)$ numerically. The resulting borderline of the eigenvalue domain is shown in \fref{figfree} (dashed lines) together with the eigenvalue distribution of the matrix ${\hat X } = {\hat C} + \rmi ({\hat S}' - {\hat \mathbb{I}})$ found by the numerical diagonalization of a set of $10^4 \times 10^4$ random matrices. At the smallest density considered $\rho \lambda_0^3 = 0.01$, the borderline found using \eref{bluerho0} is very close to \eref{domain}. At higher densities the former describes numerical results much better than \eref{domain}.

\Eref{domain} predicts a splitting of the eigenvalue domain in two parts at $\beta = 8$. The more accurate calculation using \eref{bluerho0} makes a similar prediction (see the lower right panel of \fref{figfree}). However, the eigenvalues of the matrix ${\hat X}$ do not show such a splitting and form an `inverted T' distribution on the complex plane instead. This is due to the fact that the Marchenko-Pastur law \eref{pmp} fails to describe the eigenvalue distribution of the matrix ${\hat S}'$ at $\beta > 1$ and hence the Blue function $1/(1-\beta z) + 1/z$ that we assumed for ${\hat S}'$ is not a good approximation anymore.

It is worthwhile to note that large random non-Hermitian matrices similar to our matrix ${\hat X}$ were considered previously by Haake {\it et al.} \cite{haake92} (with the help of the replica trick), Lehmann {\it et al.} \cite{lehmann95} (using the supersymmetry method) and Janik {\it et al.} \cite{janik97} (using the free probability theory). These authors studied matrices of the form ${\hat H} + \rmi c {\hat \Gamma}$, where ${\hat H}$ was an Hermitian matrix with random elements obeying Gaussian statistics, ${\hat \Gamma}$ was a Wishart random matrix [i.e. a matrix of the form \eref{nmhh}], and $c$ was a real number controlling the `degree of non-Hermiticity' of the matrix. The splitting of the domain of existence of eigenvalues in two parts was observed when $c$ was increased. This is different from our matrix ${\hat X}$ that has elements with equal variances $\beta/N$ of real and imaginary parts (hence always the same degree of non-Hermiticity) but that still exhibits the splitting of the eigenvalue domain when $\beta$ is increased.

\section{Eigenvalue distribution of the complex expc matrix}
\label{secexpc}

By analogy with the cardinal sine and cosine functions, a `cardinal complex exponent' function can be defined as $f(x) = \exp(\rmi x)/x$. The Euclidean random matrix ${\hat G}$ corresponding to this function has elements
\begin{eqnarray}
G_{ij} = f(\mathbf{r}_i - \mathbf{r}_j) =
(1 - \delta_{ij})
\frac{\exp(\rmi k_0 |\mathbf{r}_i - \mathbf{r}_j|)}{
k_0 |\mathbf{r}_i - \mathbf{r}_j|}.
\label{expc}
\end{eqnarray}
This matrix has a particular importance in the problem of wave scattering by an ensemble of $N$ point-like scatterers: indeed, as we noted already in the introduction, each element of the matrix ${\hat G}$ is a Green's function of the scalar Helmholtz equation \eref{helmholtz}.

Although the matrix ${\hat G}$ is similar to the matrix ${\hat X}$ considered in the previous section, the analytic study of its properties is much more involved. On the one hand, similarly to the eigenvalues of ${\hat X}$, the eigenvalues of ${\hat G}$ obey $\mathrm{Im} \lambda > -1$. On the other hand, correlations that arise between the real and imaginary parts of ${\hat G}$ due to the presence of the same set of points $\{ \mathbf{r}_i \}$ in both $\mathrm{Re} {\hat G} = {\hat C}$ and $\mathrm{Im} {\hat G} = {\hat S} - {\hat \mathbb{I}}$, do not permit to take full advantage of the free probability approach described in \sref{secfree}. Another way to deal with non-Hermitian matrices is to double the size of the space and to manipulate Hermitian matrices of size $2N \times 2N$ (in the `quaternion' space \cite{janik97} or in the `chiral' space \cite{feinberg97}). Due to technical difficulties, however, this approach can be readily put in practice only in certain special cases like, e.g., in the case of circularly invariant distributions $p(\lambda) = p(|\lambda|)$ \cite{feinberg06}.

\begin{figure}
\vspace{2mm}
\centering{
\includegraphics[angle=-90,width=0.99\textwidth]{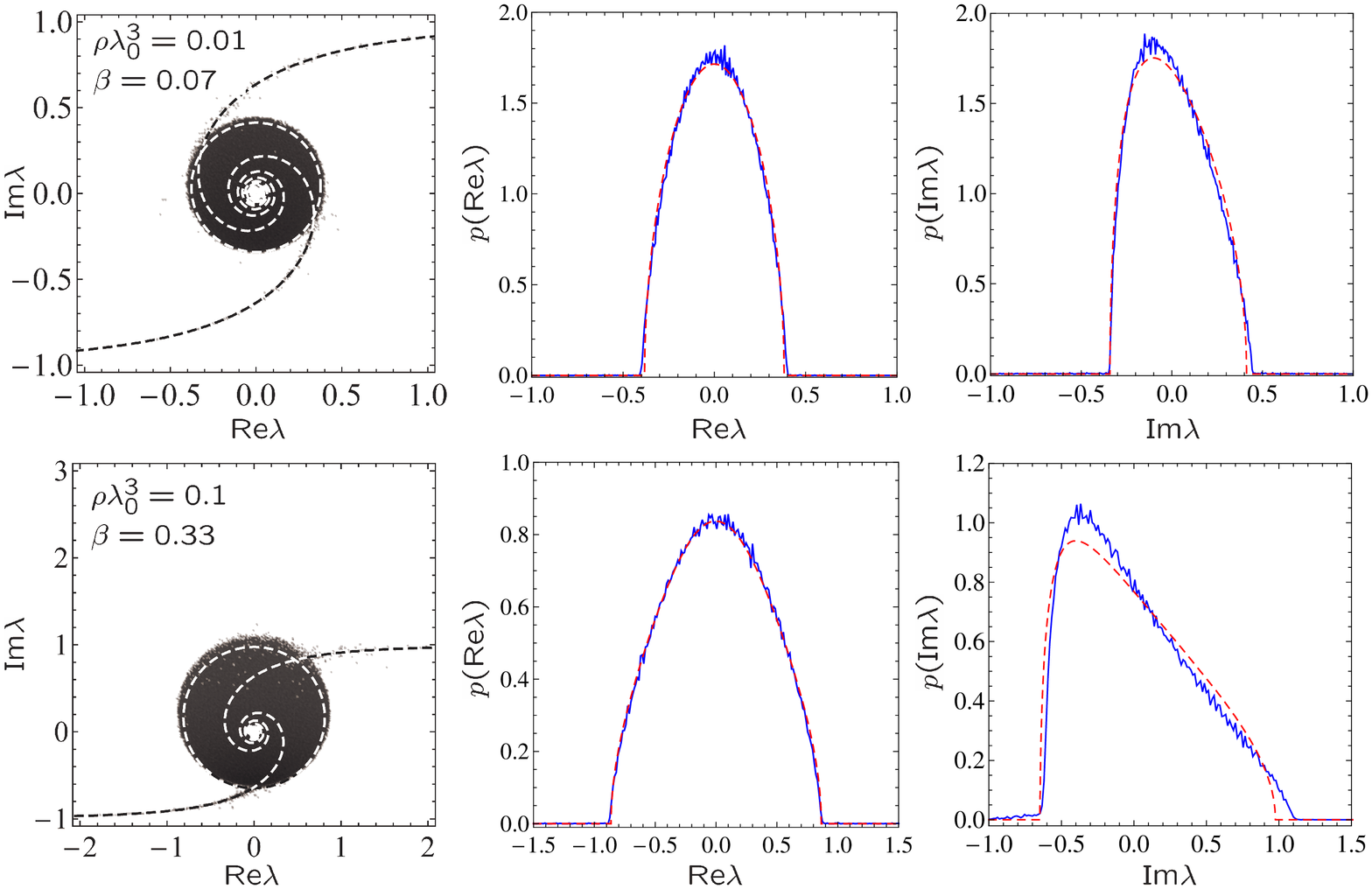}
\hspace{1.5cm}
\caption{\label{figexplow}
Left column: density plot of the logarithm of the probability density of eigenvalues $\lambda$ of a square $N \times N$ Euclidean matrix ${\hat G}$ with elements
$G_{ij} = (1-\delta_{ij}) \exp(\rmi k_0 |\mathbf{r}_i - \mathbf{r}_j|)/
k_0 |\mathbf{r}_i - \mathbf{r}_j|$ at low densities $\rho \lambda_0^3 = 0.01$ (first row) and 0.1 (second row), $\lambda_0 = 2\pi/k_0$. $N = 10^4$ points $\mathbf{r}_i$ are randomly chosen inside a 3D cube. Dashed circles are centered at $(0, \frac12 \beta)$ and have radii $\sqrt{2 \beta}$. Eigenvalues of a $2 \times 2$ matrix would lie on dashed spirals.
Central column: marginal probability density of the real part of $\lambda$ compared to our equation \eref{bluefull1} with $\beta$ replaced by $\frac12 \beta$ (dashed red line).
Right column: marginal probability density of the imaginary part of $\lambda$ compared to the Marchenko-Pastur law \eref{pmp} with $\lambda$ replaced by $\mathrm{Im} \lambda + 1$ and $\beta$ replaced by $\frac12 \beta$ (dashed red line).}}
\end{figure}

\begin{figure}
\vspace{2mm}
\centering{
\includegraphics[angle=-90,width=0.99\textwidth]{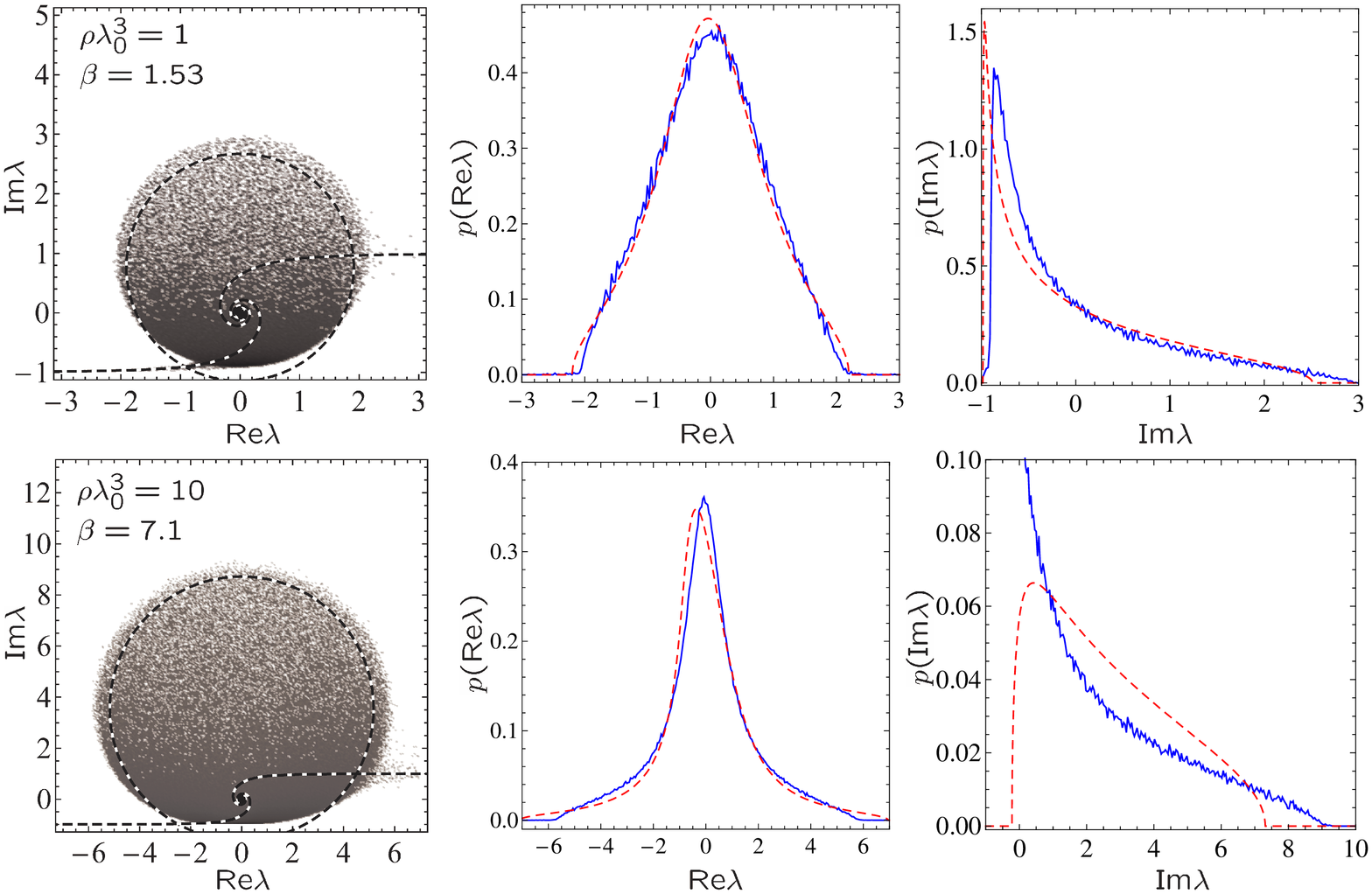}
\hspace{1.5cm}
\caption{\label{figexpinter}
Same as \fref{figexplow} but for intermediate densities $\rho \lambda_0^3 = 1$ (first row) and 10 (second row).
Dashed circles are centered at $(0, \frac12 \beta)$ and have radii $R$ given by \eref{radius}.}}
\end{figure}

\begin{figure}
\vspace{3mm}
\centering{
\includegraphics[angle=-90,width=0.99\textwidth]{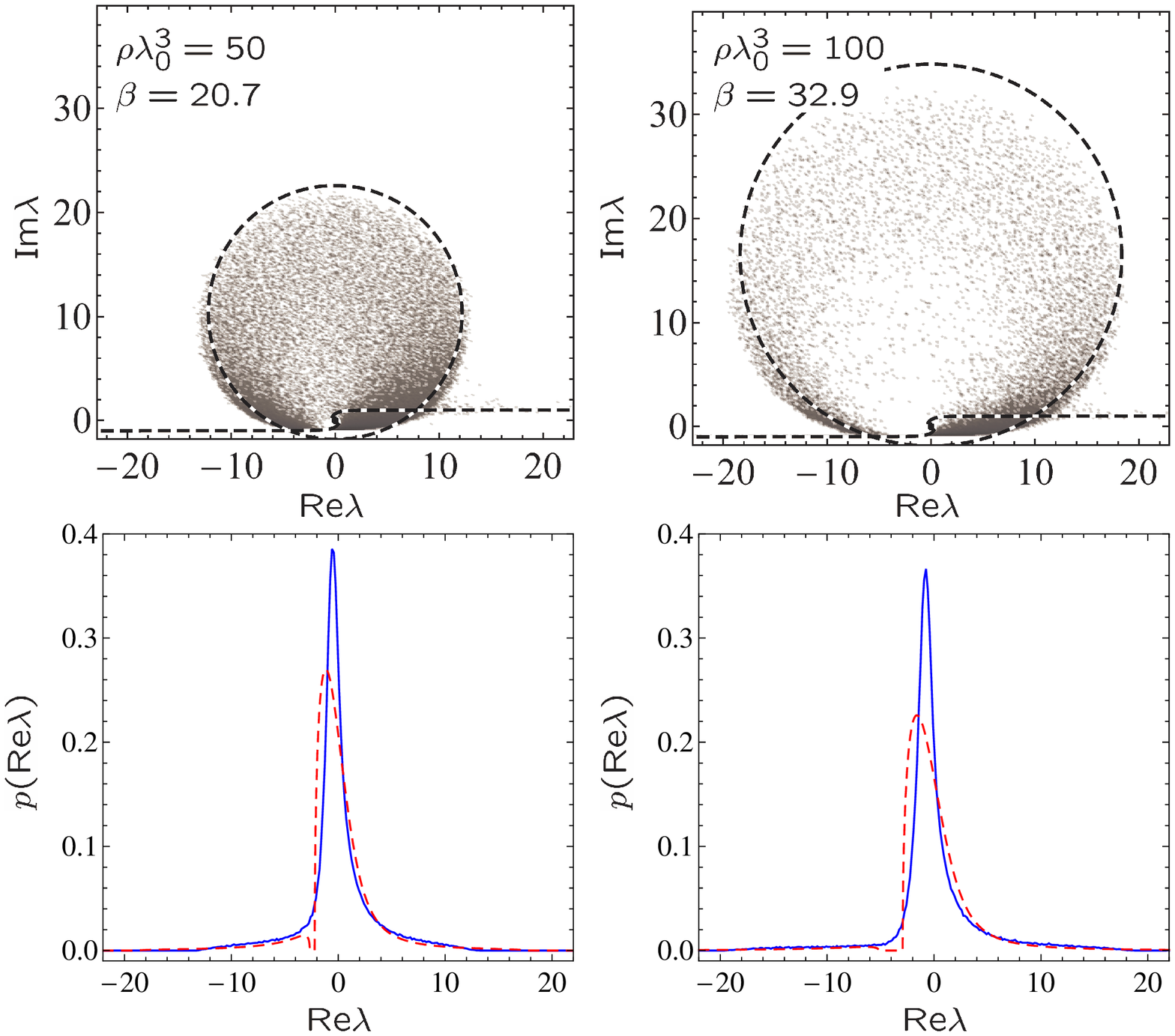}
\hspace{1.5cm}
\caption{\label{figexphigh}
Same as \fref{figexplow} and \fref{figexpinter} but for high densities $\rho \lambda_0^3 = 50$ (left column) and 100 (right column); dashed circles as in \fref{figexpinter}. Note a hole that develops on the left from $\mathrm{Re} \lambda = 0$ near the real axis and the corresponding gap in the analytic result for the marginal distributions of $\mathrm{Re} \lambda$ (dashed red lines). Marginal distributions of $\mathrm{Im} \lambda$ are not shown.}}
\end{figure}

Despite the differences between the matrices ${\hat G}$ and ${\hat X}$, a comparison of their eigenvalue distributions appears to be quite useful. Numerical calculations show that, roughly speaking, the eigenvalues of ${\hat G}$ are concentrated within a circle on the complex plane (see \fref{figexplow}, \fref{figexpinter} and \fref{figexphigh}). The same circular shape of the domain of existence of eigenvalues is characteristic for the matrix ${\hat X}$ in the limit of
$\beta \rightarrow 0$.
The radius of the circle is $\sqrt{2 \beta}$ and the position of its center is $x_0 = 0 $, $y_0 = \frac12 \beta$ as can be seen from \eref{domain} by assuming $y \ll 1$ in the denominators. To derive this result in a more rigorous way, we substitute the equation of a circle, $x = r \sqrt{2 \beta} \cos \phi$ and
$y = r \sqrt{2 \beta} \sin \phi + \frac12 h \beta$, in \eref{domain}. The l.h.s. of the resulting equation is then expanded in orders of $\beta$ and coefficients $c_{\nu}$ in front of consecutive powers of $\beta$ are analyzed. The coefficient $c_0$ in front of $\beta^0$ is zero whatever $r$ and $h$. The coefficient in front of $\beta^1$ is $c_1 = 2(r^2 - 1)$. \Eref{domain} is therefore obeyed up to the linear order in $\beta$ for $r = 1$ and any $h$. The next term is of the order $\beta^{3/2}$ and it cannot be put to zero simply by adjusting $h$ because $c_{3/2}$ depends on $\phi$:
$c_{3/2}(\phi) = \sqrt{2} \sin \phi (h + 2 \cos 2 \phi)$. We thus search for $h$ that minimizes the integral
$\int_0^{2\pi} d\phi\; c_{3/2}^2(\phi) =
2 \pi (h^2 - 2 h + 2)$. This yields $h = 1$.
The density of eigenvalues inside the circle is not homogeneous: expanding \eref{pfree} around $y = y_0$ and taking the limit $\beta \rightarrow 0$ we obtain $p(x,y) \simeq (1 - y)/2\pi \beta$.
In \fref{figexplow} we superimpose the circle of radius $\sqrt{2 \beta}$ centered at $x_0 = 0$, $y_0 = \frac12 \beta$ on the eigenvalue distribution of ${\hat G}$ for small $\beta$. The circle describes the boundary of the  eigenvalue distribution remarkably well. We thus conclude that in the limit of $\beta \ll 1$, the domains of existence of eigenvalues of the matrices ${\hat G}$ and ${\hat X}$ are very similar.
In addition to the eigenvalues inside the circle, ${\hat G}$ has eigenvalues that follow the spirals corresponding to the eigenvalues $G_{12}$ and $-G_{12}$ of a $2 \times 2$ matrix ${\hat G}$.
Interestingly, the spirals are quite robust and survive at all densities (see \fref{figexplow}, \fref{figexpinter} and \fref{figexphigh}).

Because the matrices ${\hat S}$ and ${\hat C}$ studied in previous sections represent the imaginary and real parts of the matrix ${\hat G}$, respectively, one might expect some links between the probability distributions of eigenvalues of ${\hat S}$ and ${\hat C}$ and the marginal probability distributions of the real and imaginary parts of the eigenvalues of ${\hat G}$. And indeed, we see from the central and right columns of \fref{figexplow} that the marginal probability distributions $p(\mathrm{Re} \lambda)$ and $p(\mathrm{Im} \lambda)$ are nicely described by \eref{bluefull1} and \eref{pmp}, respectively, with $\beta$ replaced by $\frac12 \beta$. This suggests an interesting interpretation of the Marchenko-Pastur law \eref{pmp}: it can be seen as a projection of a two-dimensional distribution $p(x,y)$ of complex eigenvalues $x + \rmi y$ on the imaginary axis $y$, provided that $p(x,y)$ is different from zero only inside a circle of radius $2\sqrt{\beta}$ centered at $(0, \beta)$ and that $p(x,y) \propto 1/y$ inside the circle. $p(x,y)$ being independent of $x$ and decaying monotonically with $y$ is consistent with the result that we obtained for the matrix ${\hat X}$ using the free random variable theory in the limit of $\beta$, $\rho \lambda_0^3 \rightarrow 0$.

When we increase $\beta$ but keep the density relatively low ($\rho \lambda_0^3 < 30$, see below), the cloud of eigenvalues grows but keeps its circular shape (see \fref{figexpinter}) \footnote{Because we present results at a fixed $N = 10^4$, increasing $\beta$ is achieved by increasing the density $\rho \lambda_0^3$. However, by repeating the analysis at $N = 10^3$ and $N = 5 \times 10^3$ we checked that the distributions presented in \fref{figexpinter} change only slightly when $N$ and $\rho \lambda_0^3$ are varied to keep $\beta$ constant.}.
The distribution of eigenvalues acquires an important asymmetry: the eigenvalues are `attracted' by the axis $\mathrm{Im} \lambda = -1$.
Interestingly, whereas the Marchenko-Pastur law ceases to describe the marginal distribution of $\mathrm{Im} \lambda$ when $\frac12 \beta$ becomes larger than unity, the region of validity of \eref{bluefull1} for the distribution of $\mathrm{Re} \lambda$ is wider: as we show in \fref{figexpinter}, \eref{bluefull1} continues to yield reasonable results even for $\frac12 \beta > 1$.
As can be seen from \fref{figexpinter}, even at $\frac12 \beta \gtrsim 1$ the borderline of the eigenvalues' domain is still roughly a circle. More accurate inspection reveals that this circle is still centered at $(0, \frac12 \beta)$ even for $\beta \gg 1$. It touches the line $\mathrm{Im} \lambda = -1$ that it cannot cross. Its radius is, therefore, roughly $\frac12 \beta$ and not $\sqrt{2 \beta}$ as in the limit of small $\beta$. To extrapolate between the limits of small and large $\beta$ we propose the following empirical expression for the radius $R$ of the eigenvalue domain:
\begin{eqnarray}
R^2 \approx 2 \beta + \left( \frac{\beta}{2} \right)^2.
\label{radius}
\end{eqnarray}
For $\beta \ll 1$, the second term of this equation is negligible and we recover $R = \sqrt{2 \beta}$. For the parameters of \fref{figexplow}, for example, a circle of radius $R$ given by \eref{radius} is virtually indistinguishable from the circle of radius $\sqrt{2 \beta}$ shown in the figure. At larger $\beta$ the second term in \eref{radius} starts to play a role and dominates for $\beta \gg 1$. As we show in \fref{figexpinter}, \eref{radius} gives a good idea of the part of the complex plane where the eigenvalues of the matrix ${\hat G}$ are concentrated.

At high densities $\rho \lambda_0^3 \gtrsim 30$, a `hole' appears in the eigenvalue distribution that otherwise still preserves its overall circular structure (see \fref{figexphigh}) \footnote{By repeating calculations with $N = 10^3$ and $N = 5 \times 10^3$ we found that the density $\rho \lambda_0^3$ at which the hole appears in the eigenvalue distribution is roughly independent of $\beta$.}. Interesting enough, this hole is not accompanied by any visible signatures in the marginal distributions $p(\mathrm{Re} \lambda)$ and $p(\mathrm{Im} \lambda)$. However, the analytic result \eref{bluefull1} develops a gap precisely at the same density $\rho \lambda_0^3 \approx 30$ and at the same position at which the hole appears on the complex plane. This suggests that even though \eref{bluefull1} does not provide a correct description of the marginal distribution $p(\mathrm{Re} \lambda)$ at such high densities, it still reflects some relevant properties of the distribution of complex eigenvalues $\lambda$. Note that at high densities $\rho \lambda_0^3$ the eigenvalue distribution is concentrated near the axis $\mathrm{Im} \lambda = -1$ and the parts of the distribution corresponding to $\mathrm{Im} \lambda \gg 1$ in \fref{figexphigh} are visible only thanks to the logarithmic scale of the plot.

\begin{figure}
\vspace{2mm}
\centering{
\includegraphics[angle=-90,width=0.6\textwidth]{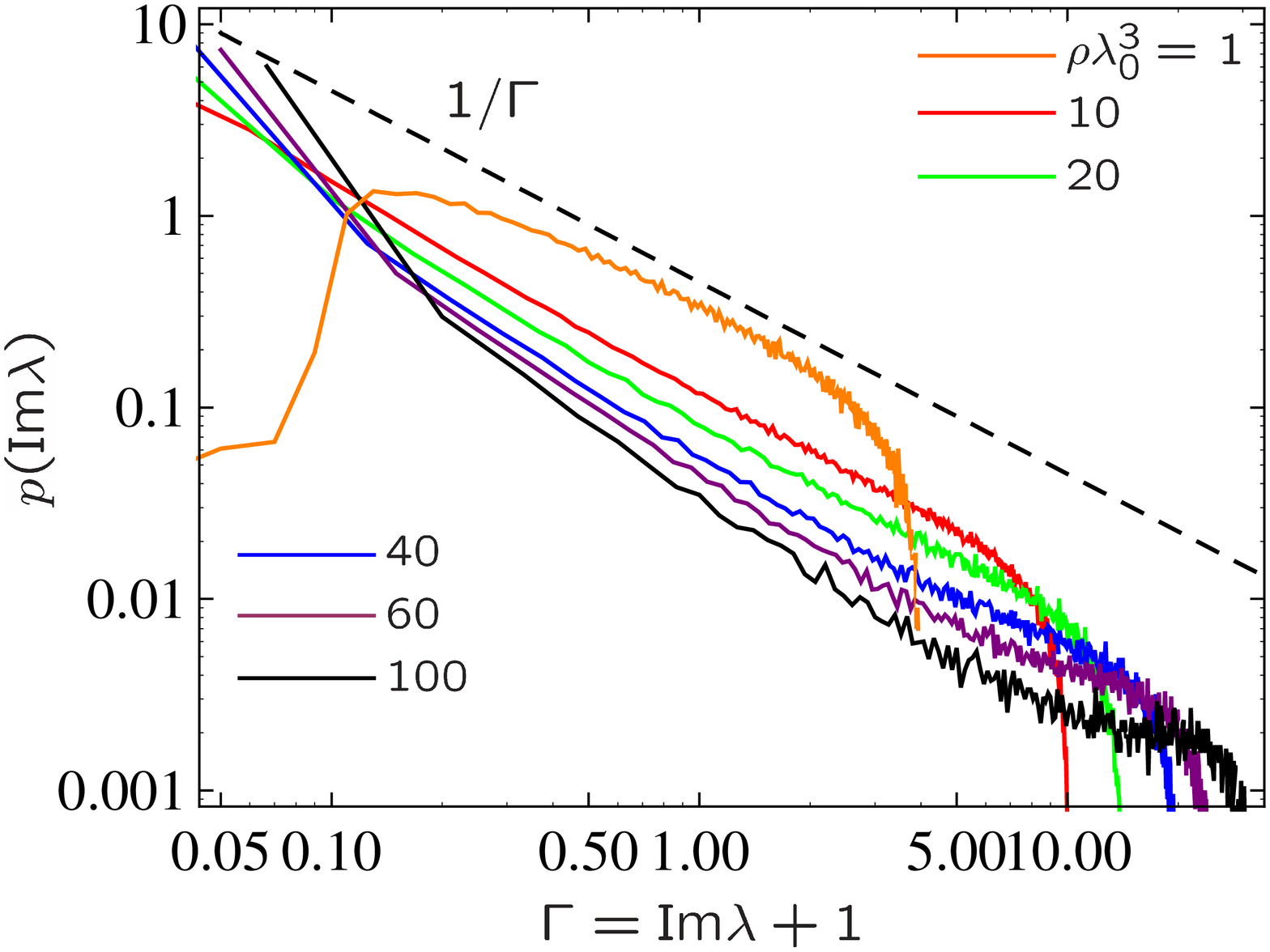}
\hspace{1.5cm}
\caption{\label{figexpim}
Marginal distribution of the imaginary part of the eigenvalues of the matrix ${\hat G}$ computed numerically at densities $\rho \lambda_0^3 = 1$, 10, 20, 40, 60 and 100 (curves from top to bottom) for $N = 10^4$ are compared with the asymptotic law $1/\Gamma$ shown by the dashed line.}}
\end{figure}

\begin{figure}
\vspace{2mm}
\centering{
\includegraphics[angle=-90,width=\textwidth]{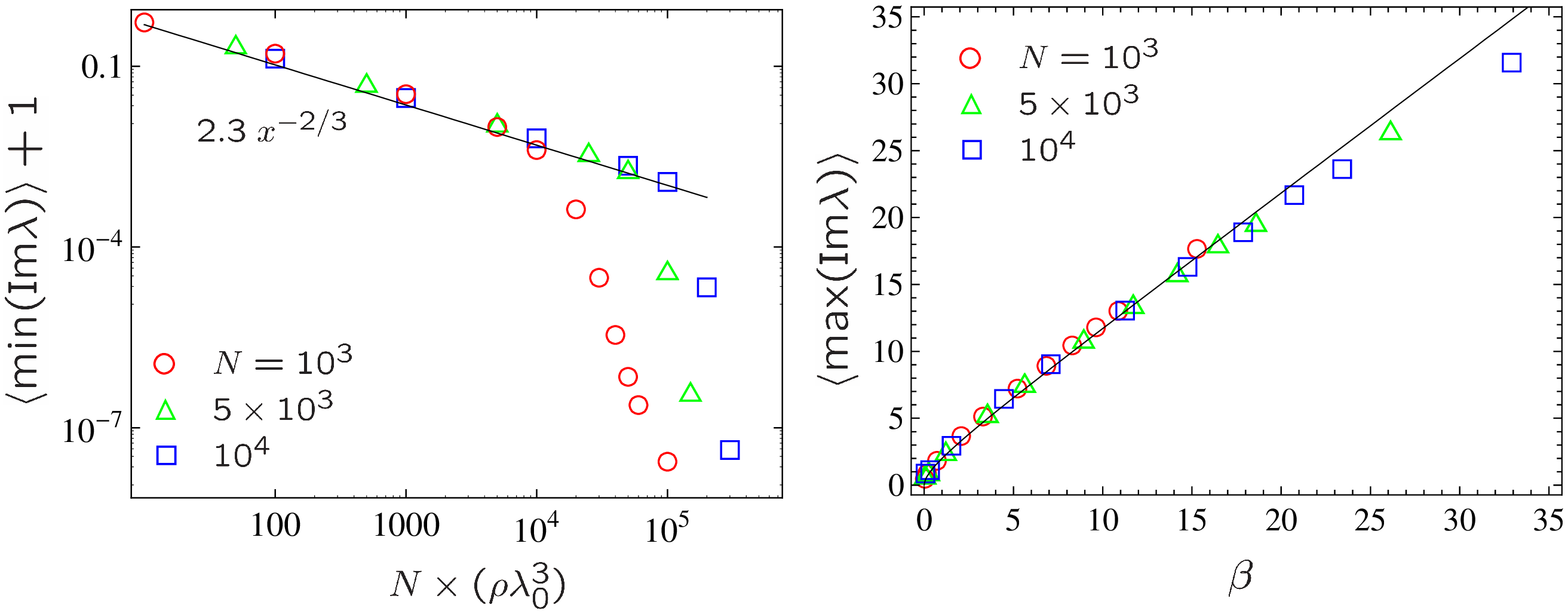}
\hspace{1.5cm}
\caption{\label{figexpminmax}
Mean minimum (left) and maximum (right) values of the imaginary part of eigenvalues $\lambda$ of the matrix ${\hat G}$ for three different matrix sizes $N$ (symbols). $\langle \mathrm{Min}(\mathrm{Im} \lambda) \rangle + 1$ is approximately $2.3 [N (\rho \lambda_0^3)]^{-2/3}$ [solid line in the left panel and \eref{imlmin}] for $\rho \lambda_0^3 \lesssim 10$ and decays faster at higher densities. $\langle \mathrm{Max}(\mathrm{Im} \lambda) \rangle$ scales with $\beta$. Different values of $\beta$ are obtained by changing the density $\rho \lambda_0^3$ from 0.01 to 100. The solid line in the right panel shows $\langle \max(\mathrm{Im} \lambda) \rangle = \frac12 \beta + R$ with $R$ given by \eref{radius}.}}
\end{figure}

Finally, we study the marginal distribution of $\mathrm{Im} \lambda$. It has been given special attention previously because, under certain assumptions, it was shown to give the distribution of `decay rates' $\Gamma = \mathrm{Im} \lambda + 1$ of quasi-modes in an open random medium \cite{rusek00,pinheiro04}. When $\frac12 \beta > 1$, $p(\mathrm{Im} \lambda)$ does not follow the Marchenko-Pastur law anymore (see \fref{figexpinter}). Based on the results of numerical simulations, Pinheiro {\it et al.} \cite{pinheiro04} claimed that at high densities $\rho \lambda_0^3$ the marginal distribution $p(\mathrm{Im} \lambda)$ exhibits a universal $1/\Gamma$ decay. Our analysis summarized in \fref{figexpim} confirms that such a decay is present, even though it seems to speed up slightly when the density is increased.
In certain applications of random matrix theory to wave propagation in random media and, in particular, in problems related to Anderson localization (see \sref{loca}) and random lasing (see \sref{lasers}), a special role is played by the eigenvalue of ${\hat G}$ that have the smallest or the largest imaginary part.
Both $\min(\mathrm{Im} \lambda)$ and $\max(\mathrm{Im} \lambda)$ are random variables. Let us first consider $\min(\mathrm{Im} \lambda)$. As can be seen from \fref{figexplow} and \fref{figexpinter}, at moderate densities $\rho \lambda_0^3 \lesssim 10$, $\min(\mathrm{Im} \lambda)$ is due to the lower spiral emerging from the `bulk' of the distribution. Eigenfunctions of ${\hat G}$ corresponding to spirals are localized on pairs of nearby points and the eigenvalues can be found by considering a $2 \times 2$ matrix ${\hat G}$. For two points at a distance $\Delta r$ we find the eigenvalues $\lambda_{1,2} = \pm \exp(\rmi k_0 \Delta r)/k_0 \Delta r$, with $\lambda_2$ corresponding to the lower spiral. The smallest values of $\mathrm{Im} \lambda$ are achieved for small distances $\Delta r$ when we can approximately write
$\mathrm{Im} \lambda_2 = -\sin(k_0 \Delta r)/k_0 \Delta r \simeq -1 + (k_0 \Delta r)^2/6$. Hence, the statistical distribution of $\min(\mathrm{Im} \lambda)$ is directly related to the statistical distribution $p(\Delta r_{\mathrm{min}})$ of the minimal distance $\Delta r_{\mathrm{min}}$ between any 2 points among $N$ points in the volume $V$. The distribution $p(\Delta r_{\mathrm{min}})$ can be constructed as follows.
Let us choose an arbitrary point $i$. The probability that another point $j$ is located in a spherical shell of radius $r$ and thickness $\rmd r$ around the first is $p_1 = 4 \pi r^2 \rmd r/V$. For $r$ to be the minimal distance $\Delta r_{\mathrm{min}}$ we have to require that all other $N-2$ points are outside the sphere of radius $r$ [probability
$p_2 = (1 - 4\pi r^3/3V)^{N-2}$] and that the distances between the remaining $(N-1)^2$ pairs of points not including the point $i$ exceed $r$ [probability
$p_3 = (1 - 4\pi r^3/3V)^{(N-1)^2}$]. The probability that $r$ is the minimum distance between any 2 points is then equal to the number of possibilities $N(N-1)$ to choose the two points $i$ and $j$, times $p_1 \times p_2 \times p_3$. The probability density is then
\begin{eqnarray}
p(\Delta r_{\mathrm{min}}) =
N(N-1) \left( \frac{4 \pi \Delta r_{\mathrm{min}}^2}{V}
\right) \left(1 - \frac{4 \pi \Delta r_{\mathrm{min}}^3}{3V}
\right)^{N(N-1)-1}.
\label{prmin}
\end{eqnarray}
This distribution is normalized to 1 if we assume that the volume $V$ is spherical (radius $R_0$) and that $\Delta r_{\mathrm{min}}$ can vary from $0$ to $R_0$.
Because
$\min(\mathrm{Im} \lambda) = -1 + k_0^2 \Delta r_{\mathrm{min}}^2/6$, its probability density is equal to
$p[\Delta r_{\mathrm{min}} = \sqrt{6 (\min(\mathrm{Im} \lambda) + 1)}/k_0] \times [\rmd \Delta r_{\mathrm{min}}/\rmd \min(\mathrm{Im} \lambda)]$. In particular, the first moment of this distribution in the limit of $N \rightarrow \infty$ is
\begin{eqnarray}
\langle \min(\mathrm{Im} \lambda) \rangle \simeq
-1 + \frac{\pi^{4/3} \Gamma(5/3)}{6^{1/3}}
\times \frac{1}{(\rho \lambda_0^3 \times N)^{2/3}}.
\label{imlmin}
\end{eqnarray}
We compare this result with numerical simulations in \fref{figexpminmax} (left panel) and find good agreement for densities $\rho \lambda_0^3 \lesssim 10$. At higher densities, $\langle \min(\mathrm{Im} \lambda) \rangle$ is smaller than predicted by \eref{imlmin}, signaling that $\min(\mathrm{Im} \lambda)$ is not dominated by the eigenvalues corresponding to eigenfunctions localized on pairs of points anymore.

Similarly to $\min(\mathrm{Im} \lambda)$, $\max(\mathrm{Im} \lambda)$ is dominated by the second spiral branch of the eigenvalue distribution for $\beta \lesssim 0.3$ (see \fref{figexplow}). At larger $\beta$, $\max(\mathrm{Im} \lambda)$ belongs to the bulk of the eigenvalue distribution (see \fref{figexpinter} and \fref{figexphigh}). As follows from our analysis, the distribution of complex eigenvalues $\lambda$ of the matrix ${\hat G}$ occupies a circular domain of radius $R$ given by \eref{radius}, centered at $\frac12 \beta$. It follows then that
\begin{eqnarray}
\langle \max(\mathrm{Im} \lambda) \rangle \approx \frac12 \beta + R.
\label{imlmax}
\end{eqnarray}
And indeed, this approximate expression describes numerical results quite reasonably (see the solid line in the right panel of \fref{figexpminmax}), even though a closer inspection reveals that it overestimates $\langle \max(\mathrm{Im} \lambda) \rangle$ at large $\beta$. Further work is needed to find a more accurate expression for $\langle \max(\mathrm{Im} \lambda) \rangle$.

\section{Applications}
\label{appl}

\noindent
We have already mentioned in the introduction that the Euclidean random matrices ${\hat S}$, ${\hat C}$ and ${\hat G}$ studied in this paper are encountered in several physical problems. In this section we briefly discuss a number of such problems and show how our results can help to advance their understanding.

\subsection{Cooperative emission of large atomic clouds}
\label{coop}

An interesting problem of modern quantum optics is the one in which a single photon is stored in a cloud of (cold) atoms. One studies the properties (frequency, direction of propagation, etc.) of the photon re-emitted by the cloud at a later time \cite{svid08,scully09,svid09,svid10}. For $N$ two-level atoms (excited state $a$, ground state $b$) located at random points $\mathbf{r}_i$, $i = 1, \ldots, N$, the state of the system at a time $t$ can be written as \cite{svid10}
\begin{eqnarray}
\Psi(t) &=& \sum\limits_{j=1}^{N} \beta_j(t)
| b_1 b_2 \cdots a_j \cdots b_N \rangle |0\rangle
+
\sum\limits_{\mathbf{k}} \gamma_{\mathbf{k}}(t)
| b_1 b_2 \cdots b_N \rangle |1_{\mathbf{k}}\rangle
\nonumber \\
&+&
\sum\limits_{m < n} \sum\limits_{\mathbf{k}}
\alpha_{mn, \mathbf{k}}
| b_1 b_2 \cdots a_m \cdots a_n \cdots b_N \rangle
|1_{\mathbf{k}}\rangle.
\label{state1}
\end{eqnarray}
Here the first sum corresponds to the superposition of states with one atom (atom $j$) in the excited state, all other atoms in the ground state, and zero photons. The second sum corresponds to the states in which all atoms are in the ground state, while there is a photon in the mode ${\mathbf{k}}$. Finally, the last sum describes states with atoms $m$ and $n$ in the excited state and one virtual photon with `negative' energy.

The evolution equation for the vector $\bbeta(t) = \{\beta_j(t)\}$ reads \cite{svid09,svid10}:
\begin{eqnarray}
\dot{\bbeta}(t) = -\Gamma_0 \bbeta(t) + \rmi \Gamma_0
{\hat G} \bbeta(t),
\label{betadot}
\end{eqnarray}
where $\Gamma_0$ is the spontaneous decay rate of a single atom and the matrix ${\hat G}$ is defined by \eref{expc}. According to this equation, a system prepared in the eigenstate described by a vector $\bbeta(0)$ decays with a rate $\Gamma_0(1 + \mathrm{Im} \lambda)$ and experiences a frequency shift $-\Gamma_0 \mathrm{Re} \lambda$, where $\lambda$ is an eigenvalue of the matrix ${\hat G}$. Both the decay rate and the frequency shift were studied in \cite{svid09,svid10} in the limit of a very dense atomic cloud ($\rho \lambda_0^3 \rightarrow \infty$), when the summation $[{\hat G} \bbeta(t)]_{j} = \sum_{m=1}^{N} G_{im} \beta_m(t)$ can be replaced by integration in the last term on the r.h.s. of \eref{betadot}. The authors also discussed a useful approximation in which the real part of the matrix ${\hat G}$ is neglected and ${\hat G}$ is replaced by $\rmi {\hat S}$ in \eref{betadot}.

Although the results of \cite{svid09,svid10} are very interesting, atomic clouds of moderate density $\rho \lambda_0^3 \lesssim 1$ are readily created in modern laboratories (see, e.g., \cite{labeyrie99,labeyrie03}). It is therefore important to extend the analysis of \cite{svid09,svid10} to such dilute atomic clouds. The present work provides, in fact, such an extension:
the distribution of dimensionless decay rates $\Gamma = 1 + \mathrm{Im} \lambda$ is given by the Marchenko-Pastur law \eref{pmp} with $\beta$ replaced by $\frac12 \beta$ and the distribution of dimensionless frequency shifts $\Omega = -\mathrm{Re} \lambda$ follows from the analysis of \sref{seccosc} (see also \fref{figexplow}, \fref{figexpinter}, \fref{figexphigh}). It is important to realize that replacing summation by integration in the last term on the r.h.s. of \eref{betadot} performed in \cite{svid09,svid10} is equivalent to averaging this equation over all possible configurations $\{ \mathbf{r}_i \}$ of atoms. It leads, therefore, to the neglect of the statistical nature of the initial problem. In contrast, our treatment does not rely on such an averaging and fully accounts for large fluctuations of eigenvalues, typical for situations when light is scattered in a strongly disordered environment. As a consequence, the authors of \cite{svid09,svid10} find deterministic eigenvalues $\lambda_n$, whereas we work with the probability distribution $p(\lambda)$. Our results are consistent with those of \cite{svid09,svid10} in the limit of $\rho \lambda_0^3 \rightarrow \infty$ and provide a generalization of some of them. For example, the authors of \cite{svid09,svid10} predict that for $\rho \lambda_0^3 \rightarrow \infty$, the fastest decay rate $\Gamma_{\mathrm{max}}$ is of the order of $\beta$. Our study suggests that the dependence of $\Gamma_{\mathrm{max}}$ on $\beta$ (and not on the density $\rho \lambda_0^3$) is a general property valid at any density (see the right panel of \fref{figexpminmax}) as well as it yields a more precise relation between $\Gamma_{\mathrm{max}} = 1 + \max(\mathrm{Im} \lambda)$ and $\beta$ [see \eref{imlmax}].

\subsection{Anderson localization in an open medium}
\label{loca}

The phenomenon of Anderson localization is common for all waves in random media \cite{anderson58,lagendijk09,alspecial10}. It consists in a transition from extended (over the whole available sample volume) to exponentially localized eigenstates of a wave (or Schr\"{o}dinger) equation with a randomly fluctuating dielectric constant (or potential), at a sufficiently strong randomness. A paradigm system in which Anderson localization can be studied for classical waves is a random arrangement of $N$ identical point-like scatterers in a volume $V$. In such an open system of finite size the wave energy can leak to the outside and one expects Anderson localization to have an impact on decay of physical observables (such as, e.g., the intensity of the wave emerging from the random system). Given a simple model for scatterers, the relevant decay rates are related to the imaginary part of the eigenvalues $\lambda$ of the non-Hermitian matrix $\hat G$ \cite{rusek00}.

Several authors studied the distribution of dimensionless decay rates $\Gamma= \mathrm{Im} \lambda + 1$ in open random media and, in particular, promoted the idea of using its probability distribution $p(\Gamma)$ as a criterion for Anderson localization \cite{pinheiro04,kottos05}. More precisely, $p(\Gamma)$ is expected to decay as  $1/\Gamma$ in the localized regime. Our numerical results also exhibit such a behavior (see \fref{figexpim}), but we cannot claim any relation between it and Anderson localization. Indeed, a careful inspection of our results shows that $p(\Gamma)$ starts to exhibit $1/\Gamma$ behavior right after the criterion $\frac12 \beta < 1$ breaks down.
On the one hand, for resonant point-like scatterers the mean free path can be estimated in the independent scattering approximation as $\ell = 1/\rho \sigma = k_0^2/4 \pi \rho$ with the resonant scattering cross-section $\sigma = 4\pi/k_0^2$. The criterion $\frac12 \beta = 1$ corresponds then to a condition for the optical thickness $L/\ell \simeq 9$. On the other hand, Anderson localization is expected to take place for $k_0 \ell \simeq 1$ (Ioffe-Regel criterion \cite{lagendijk09}) which can be rewritten as $\rho \lambda_0^3 \simeq 20$. We thus see that the condition required to observe $1/\Gamma$ decay of $p(\Gamma)$ ($L/\ell \gtrsim 9$) does not seem to agree with the one expected for the Anderson localization ($\rho \lambda_0^3 \gtrsim 20$).


The results that we obtained in the present paper suggest another way of using statistics of eigenvalues of ${\hat G}$ to look at the transition from weak to strong scattering and eventually to Anderson localization. First, instead of studying the imaginary part of $\lambda$ one can study its real part. At low density $\rho \lambda_0^3 \ll 1$ the distribution $p(\mathrm{Re} \lambda)$ exhibits a transition from the Wigner semi-circle law for $\beta \sim L/\ell \ll 1$ (see \fref{figexplow}) to the Cauchy distribution for $\beta \sim L/\ell \gg 1$ (see \fref{figexpinter}). This transition can be seen as a signature of the change of regime of wave scattering from single (for $L/\ell \ll 1$) to multiple (for $L/\ell \gg 1$) scattering. Second, an important modification of $p(\lambda)$ that takes place when the density of scatterers is increased is the appearance of a hole in the distribution that otherwise occupies a circular domain on the complex plane. The condition $\rho \lambda_0^3 \simeq 30$ for the appearance of the hole is remarkably close to the condition $\rho \lambda_0^3 \simeq 20$ expected for the Anderson localization transition in the independent scattering approximation. A highly speculative conjecture might be that a link exists between the hole in $p(\lambda)$ on the complex plane and Anderson localization of waves in an ensemble of point-like scatterers. Further work is required to prove or to refute this conjecture.

\subsection{Random lasers and optical instabilities}
\label{lasers}

The eigenvalues of the matrix ${\hat G}$ that have the smallest imaginary part play a particularly important role for understanding of very interesting optical systems called `random lasers'. Random laser is a laser that have no external cavity and in which the feedback is provided by the multiple scattering of light \cite{cao03,cao05,wiersma08}. One of the minimal models to study random lasing is an ensemble of point-like scatterers (`atoms') randomly distributed in a volume $V = L^3$ filled with some continuous amplifying medium that provides a constant amplification rate $\Gamma_{\mathrm{ampl}}$. Lasing starts when $\Gamma_{\mathrm{ampl}}$ becomes larger than the minimum loss rate $\Gamma_{\mathrm{min}} = 1 + \mathrm{min}({\mathrm{Im}} \lambda)$. Therefore, the average value of $\mathrm{min}({\mathrm{Im}} \lambda)$ defines the average random laser threshold: $\langle \Gamma_{\mathrm{ampl}}^{\mathrm{th}} \rangle = 1 + \langle \mathrm{min}({\mathrm{Im}} \lambda) \rangle$. Pinheiro and Sampaio \cite{pinheiro06} studied this latter quantity numerically and found a scaling law
\begin{eqnarray}
1 + \langle \mathrm{min}({\mathrm{Im}} \lambda) \rangle \propto \frac{1}{N^{2/3} (\rho \lambda_0^3)^{4/3}}.
\label{pinheiro}
\end{eqnarray}
They provided a simple interpretation of this result in terms of the diffusion theory of light scattering.

The eigenvalues that have the smallest imaginary part also define the threshold for dynamic instabilities in nonlinear random media. In particular, a random arrangement of point-like {\it nonlinear\/} scatterers with an intensity-dependent scattering matrix
$t(I) = (-2 \pi \rmi/k_0)[\exp(\rmi \alpha I) + 1]$ was considered by Gr\'{e}maud and Wellens \cite{gremaud10}. Here $I$ is the intensity of light on the scatterer. It was shown that stationary, time-independent solutions lose their stability and the system starts to exhibit complex, spontaneous dynamic behavior when the nonlinear coefficient $\alpha$ exceeds a critical value $\alpha_{\mathrm{inst}}$. The average value of the instability threshold was found to scale as $\langle \alpha_{\mathrm{inst}} \rangle \propto [1 + \langle \mathrm{min}({\mathrm{Im}} \lambda) \rangle]^{3/2}$, with
\begin{eqnarray}
1 + \langle \mathrm{min}({\mathrm{Im}} \lambda) \rangle \propto \frac{1}{(N \times \rho \lambda_0^3)^{2/3}}.
\label{gremaud}
\end{eqnarray}
This result can be explained by considering the spiral branches of the statistical distribution of $\lambda$ on the complex plane (see the dashed spirals in \fref{figexplow}, \fref{figexpinter} and \fref{figexphigh}); it is not related to the diffusion of light in the bulk of the random sample but originates from sub-radiant states localized on pairs of mutually close scatterers \cite{gremaud10}.

As follows from the aforesaid, the results \eref{pinheiro} and \eref{gremaud} that are supposed to coincide, not only differ by a factor $(\rho \lambda_0^3)^{-2/3}$ but they are given different physical interpretations as well. The analysis that we performed in this work allows us to resolve this controversy and to identify the result \eref{gremaud} of Gr\'{e}maud and Wellens \cite{gremaud10} as the correct one. Moreover, not only we are able to derive an analytical expression \eref{imlmin} that agrees with \eref{gremaud} and contains the precise numerical coefficient, but also the full distribution function of $\min(\mathrm{Im} \lambda)$ --- and hence of the instability threshold $\alpha_{\mathrm{inst}}$ --- follows from our result \eref{prmin}.

A random laser different from that considered in \cite{pinheiro06} is the one in which the amplification is provided by the point scatterers themselves and not by the medium in between them. Amplification and scattering in such a system are not independent anymore and cannot be tuned at will. The eigenvalues $\lambda$ of ${\hat G}$  governing the laser threshold will now depend on the specific amplification scheme. Curiously, for the simplest physical pumping mechanism we can think of (incoherent pump), the laser threshold will be determined by the eigenvalues having the {\em largest\/} imaginary part \cite{goetschy10}. This provides a direct physical application for the results that we show in the right panel of \fref{figexpminmax} and will be discussed in detail elsewhere \cite{goetschy10}.

\section{Conclusion}
\label{concl}

In this work we studied eigenvalue distributions of certain Euclidean random matrices that appear in the context of wave propagation in random media. In particular, we considered large $N \times N$ real symmetric matrices ${\hat S}$ and ${\hat C}$ with elements $S_{ij} = \sin(k_0 |\mathbf{r}_i - \mathbf{r}_j|)/k_0 |\mathbf{r}_i - \mathbf{r}_j|$ and
$C_{ij} = (1- \delta_{ij}) \cos(k_0 |\mathbf{r}_i - \mathbf{r}_j|)/k_0 |\mathbf{r}_i - \mathbf{r}_j|$, respectively, as well as the non-Hermitian matrix
${\hat G} = {\hat C} + \rmi ({\hat S} - {\hat \mathbb{I}})$. $N$ points $\mathbf{r}_i$ were chosen randomly in a three-dimensional cube of side $L$ with density $\rho = N/L^3$. For the three random matrices under study, the two important parameters of the eigenvalue distributions $p(\lambda)$ are $\beta = 2.8 N/(k_0 L)^2$ and the number of points per wavelength cube $\rho \lambda_0^3$. $\beta$ is equal to the variance of eigenvalues $\lambda$ of both ${\hat S}$ and ${\hat C}$ in the limit of $k_0 L \rightarrow \infty$.

In the low-density limit $\rho \lambda_0^3 \ll 1$ and for $\beta < 1$, the distributions of eigenvalues of Hermitian matrices ${\hat S}$ and ${\hat C}$ are parameterized uniquely by $\beta$: the distribution of eigenvalues of ${\hat S}$ is given by the Marchenko-Pastur law \eref{pmp}, whereas the distribution of eigenvalues of ${\hat C}$ can be deduced from the Blue function \eref{bluerho0} that we derived in this paper. For $\beta > 1$ the Marchenko-Pastur does not apply to ${\hat S}$ anymore, but out equation \eref{bluerho0} still works for ${\hat C}$ as long as $\rho \lambda_0^3$ is small enough. As $\beta$ increases, \eref{bluerho0} describes a transition from the Wigner semi-circle law (at $\beta \ll 1$) to the Cauchy distribution (at $\beta \gg 1$). At high densities $\rho \lambda_0^3 > 1$ the more complete expression \eref{bluefull1} that we derived for the Blue function of the matrix ${\hat C}$ applies. It is in good agreement with numerical simulations until $\rho \lambda_0^3 \approx 30$ where it predicts the appearance of a gap in $p(\lambda)$, which is not observed in the numerical data.

The eigenvalue distribution of the non-Hermitian matrix ${\hat G}$ has a circular structure on the complex plane. At $\beta \ll 1$, the eigenvalues are confined to a circle of radius $\sqrt{2 \beta}$ centered at $(0, \frac12 \beta)$. At larger $\beta$, the distribution becomes strongly asymmetric, with much stronger weight of eigenvalues with imaginary parts close to $-1$. Our numerical results show that the domain of existence of eigenvalues is still approximately a circle centered at $(0, \frac12 \beta)$. We proposed an empirical expression for its radius $R^2 \approx 2 \beta + (\frac12 \beta)^2$. At high densities $\rho \lambda_0^3 > 30$, a hole appears in the distribution $p(\lambda)$ on the complex plane. The density at which the hole appears seems to be roughly independent of $\beta$.
The marginal probability distribution of $\mathrm{Re} \lambda$
is described by our equation \eref{bluefull1} at all $\beta$, provided that $\rho \lambda_0^3 < 30$. The marginal distribution of $\mathrm{Im} \lambda$ follows the Marchenko-Pastur law \eref{pmp} for $\frac12 \beta < 1$ and decays as $1/(\mathrm{Im} \lambda + 1)$ at larger $\beta$.

Finally, we studied a model matrix ${\hat X} = {\hat C} + \rmi ({\hat S}' - {\hat \mathbb{I}})$ in which two independent ensembles of points $\{ \mathbf{r}_i \}$ and $\{ \mathbf{r}_i' \}$ were used to generate matrices ${\hat C}$ and ${\hat S}'$. The matrices ${\hat C}$ and ${\hat S}'$ are asymptotically free and the distribution of eigenvalues of ${\hat X}$ at $\beta < 1$ can be found using the approach developed by Jarosz and Nowak \cite{jarosz06} based on the theory of free random variables. The distribution of eigenvalues shows an interesting transition from a circular shape at $\beta \ll 1$ to a triangular shape at $\beta \sim 1$, and then to an `inverted T' shape for $\beta \gg 1$.

\section*{Acknowledgements}
This work was supported by the French ANR (Project No. 06-BLAN-0096 CAROL).

\end{document}